\documentclass[acmsmall]{acmart}

\acmJournal{PACMPL}
\acmVolume{1}
\acmNumber{OOPSLA} 
\acmArticle{1}
\acmYear{2020}
\acmMonth{1}
\acmDOI{} 
\startPage{1}

\setcopyright{none}

\usepackage{caption}
\usepackage{subcaption}
\usepackage{proof}
\usepackage[ruled,vlined]{algorithm} 
\usepackage{color}

\newcommand{\tab}		{\qquad}

\newcommand{\term}[1]	{\emph{\textcolor{teal}{#1}}}

\newcommand{\semantics}[1] {[\![ #1 ]\!]}
\newcommand{\typerel}[2] {#1 : #2}

\newcommand{\BD}{\begin{definition}}
\newcommand{\ED}{\end{definition}}

\DeclareMathOperator{\sat}{sat}

\newcommand{\TT}{\text{True}}
\newcommand{\FF}{\text{False}}

\newcommand{\true}  {\texttt{True}}
\newcommand{\false} {\texttt{False}}

\newcommand{\kw}[1]{\mathtt{#1}}		

\newcommand{\RQ}[1]{\textbf{RQ#1}}

\newcommand{\code}[1]{\texttt{#1}}
\DeclareMathOperator{\Prop}{Prop}
\DeclareMathOperator{\Conf}{Conf}

\newcommand{\pc}[1]     {pc_{#1}}
\newcommand{\SPL}       {\mathcal{L}}
\newcommand{\feat}[1]{\bm{#1}} 
\newcommand{\featset}{F}
\newcommand{\featmodel}{\Phi}
\newcommand{\domainmodel}{D}
\newcommand{\pcmap}{\phi}
\newcommand{\config}{\rho}

\newcommand{\indexProd}[2]{#1|_{#2}}

\renewcommand{\ifdef}{\texttt{\#ifdef}}
\newcommand{\cppendif}{\texttt{\#endif}}

\usepackage{booktabs}   
\usepackage{subcaption} 

\usepackage{graphicx}
\usepackage[utf8]{inputenc}
\usepackage[T1]{fontenc}
\usepackage{listings}
\usepackage{color}
\usepackage{algpseudocode}
\usepackage{stmaryrd}
\usepackage{bussproofs}
\usepackage{amsmath}
\usepackage{amsfonts}
\usepackage{float}
\usepackage{multicol}
\usepackage{multirow}
\usepackage{bm}
\usepackage{soul}
\usepackage{tabularx}
\usepackage{tikz-cd}

\DeclareMathOperator{\unsat}{unsat}

\newcommand{\tm}[1]		{\emph{#1}}

\newcommand{\type}[1]		{#1}

\newcommand{\F}		{F}

\renewcommand{\L}		{\mathcal{L}}

\newcommand{\prog}		{G}

\begin{document}

\title{Automatic and Efficient Variability-Aware Lifting of Functional Programs}
%
%


\author{Ramy Shahin}
\affiliation{
	\department{Computer Science Department}              
	\institution{University of Toronto}            
	\city{Toronto}
	\country{Canada}                    
}
\email{rshahin@cs.toronto.edu}          

\author{Marsha Chechik}
\affiliation{
	\department{Computer Science Department}              
	\institution{University of Toronto}            
	\city{Toronto}
	\country{Canada}                    
}
\email{chechik@cs.toronto.edu}          





\begin{abstract}
  A software analysis is a computer program that takes some representation of a software product as input and produces some useful information about that product as output.
  A software product line encompasses \emph{many} software product variants, and thus existing analyses can be applied to each of the product variations individually, but not to the entire product line as a whole. Enumerating all product variants and analyzing them one by one is usually intractable due to the combinatorial explosion of the number of product variants with respect to product line features. Several software analyses (e.g., type checkers, model checkers, data flow analyses) have been redesigned/re-implemented to support variability. This usually requires a lot of time and effort, and the variability-aware version of the analysis might have new errors/bugs that do not exist in the original one.
   
 Given an analysis program written in a functional language based on PCF, in this paper we present two approaches to
  transforming (lifting) it into a semantically equivalent variability-aware analysis. A light-weight approach (referred to as \emph{shallow lifting}) wraps the analysis program into a variability-aware version, exploring all combinations of its input arguments. Deep lifting, on the other hand, is a program rewriting mechanism where the syntactic constructs of the input program are rewritten into their variability-aware counterparts. Compositionally this results in an efficient program semantically equivalent to the input program, modulo variability.
 We present the correctness criteria for functional program lifting, together with correctness proof sketches of our program transformations. We evaluate our approach on a set of program analyses applied to the BusyBox C-language product line.
\end{abstract}

\ccsdesc[300]{Software and its engineering~Automated static analysis}
\ccsdesc[300]{Software and its engineering~Software design techniques}

\keywords{Software Product Lines, PCF, Program Rewriting, Lifting, Variability-aware Programming}

\maketitle

\newcommand{\FA}{\feat{A}}
\newcommand{\FB}{\feat{B}}
\newcommand{\FC}{\feat{C}}
\newcommand{\spl}{\mathcal{L}}
\newcommand{\ap}{G}
\newcommand{\lap}{G$^\uparrow$}
\newcommand{\lift}[1]{#1^\uparrow}

\section{Introduction}
\label{sec:intro}

%

A \term{Software Product Line (SPL)} is a family of software products, developed together from a common set of artifacts. The unit of variability in an SPL is a \term{feature}, which can be either present or absent in a given product. A \term{product} is thus considered a variant, which can be generated from the SPL given a \term{feature configuration}. In \term{annotative} SPLs, feature-specific artifacts are annotated using feature expressions.
For example, a commonly used approach for annotating the source code of an SPL is C Pre-Processor (CPP) macro definitions and conditional compilation. Features are denoted by pre-processor macro definitions, and code segments are annotated by \ifdef~blocks. Feature configurations are defined at build-time by passing macro definitions corresponding to the selected features to the CPP, which excludes the code of unselected features from being built.

Product Line Engineering (PLE) is considered much more economical in terms of time and effort compared to maintaining each of the different product variants separately~\cite{Clements:2001}. For this reason, SPLs are used in many software domains that require a high degree of variability and configurability, including operating systems, embedded systems and compilers~\cite{Pohl:2005, Liebig:2010}. For example, the Linux operating system kernel is an SPL with more than 10,000 features~\cite{Nadi:2014}.


A \term{Software Analysis Tool} is a program that takes software artifacts (e.g., source code) as input, and calculates some useful information about it. Examples include syntax checkers (parsers), type checkers, software model checkers and other bug finding tools. While such tools and PLE are very widely used individually, it is challenging to use them together. In particular, given a software analysis tool \ap~and an SPL $\spl$, we would like to be able to apply \ap~to all the product variants of $\spl$ \emph{simultaneously}, without having to enumerate the potentially exponential number of product variants and applying \ap~to each. However, since \ap~is designed to take the artifacts of a single product as input, this is not generally possible.

To get around the intractability of analyzing each possible product, some product sampling techniques have been suggested~\cite{Apel:2013,Liebig:2013}. Sampling tackles the combinatorial complexity of the problem by giving up completeness. The selected samples typically cover only a subset of possible feature interactions, which might miss many significant behaviors involving several interacting features. 

Alternatively, some analyses have been rewritten to support full SPL analysis. These include syntax checkers (parsers)~\cite{Kastner:2011,Gazzillo:2012}, type checkers~\cite{Kastner:2012}, and model checkers~\cite{Classen:2013}. Although the high level algorithm is the same as that of the corresponding single product analysis, re-developing each of these analyses from scratch is tedious, time consuming and potentially error prone. Other approaches tackle families of related analyses (e.g., data-flow analyses~\cite{Bodden:2013} and abstract interpretation~\cite{Midtgaard:2015} instead of an individual analysis. This generalizes to more analyses, given they belong to the supported family. 

A variability-aware analysis has to preserve the semantics of the original analysis, while applying it simultaneously to inputs from different product variants. At the same time, it should leverage the commonalities across the different variants as much as possible to minimize the computational overhead, and maximize the sharing of computed intermediate values. In theory the principles and techniques used in developing variability-aware analyses are the same regardless of the analysis itself. This raises the question of whether we can come up with a \emph{generic} approach to \emph{automatically} turn an \emph{arbitrary} analysis program into its equivalent variability-aware analysis. This has been accomplished for Datalog-based analyses~\cite{Shahin:2019, Shahin:2020}). Yet, Datalog is less expressive than Turing-complete languages, and many useful analyses cannot be encoded as Datalog rules. For example, program analyses that compute monotone functions over lattices cannot be expressed in plain Datalog, and require functional programming extensions (e.g., Flix~\cite{Madsen:2016}, Datafun~\cite{Arntzenius:2016}). 

In this paper, we address the problem of lifting arbitrary software analyses written in Turing-complete languages. We define a generic approach to automatically rewrite an analysis program \ap~into program \lap, where \lap~takes an SPL as input. The output of \lap~is equivalent to the collective results of applying \ap~to every product variant in the SPL. We refer to \lap~as the \term{lifted} version of \ap.

%
\lstset{
	morekeywords={ifdef, endif}
}

\setlength{\columnseprule}{1pt}
\setlength{\columnsep}{1cm}

\begin{figure}[t]
\scalebox{0.75}{}
\begin{mbox}{}	
\vspace{-0.1in}
\begin{multicols}{2}
\begin{lstlisting}[language=c,numbers=left,basicstyle=\small,xleftmargin=10pt]
int foo

#ifdef A           // A
  (int a) {
#ifdef B           // A /\ B
    return a * 2;
#else              // A /\ !B
    return a * 3 + 1;
#endif

#else              // !A
  (int a, int b) {
    return (a  
#ifdef B           // !A /\ B
	+
#else              // !A /\ !B
	* 2 +
#endif
	b);
#endif

}
\end{lstlisting}

\columnbreak

\begin{lstlisting}[language=c]
// A /\ B
int foo (int a) {
    return a * 2;
}

// A /\ !B
int foo (int a) {
    return a * 3 + 1;
}

// !A /\ B
int foo (int a, int b) {
    return (a + b);
}

// !A /\ !B
int foo (int a, int b) {
    return (a * 2 + b);
}
\end{lstlisting}
\end{multicols}
\end{mbox}
\vspace{-0.1in}
\caption{Example of an annotative product line, with the SPL on the left-hand side, and the four products that can be generated from it on the right-hand side.}
\label{fig:annotative}
\vspace{-0.1in}
\end{figure}




\vskip 0.1in

\noindent
{\bf Motivating Example.}
Fig.~\ref{fig:annotative} shows a simple example of an SPL written in the C language, where C Pre-Processor (CPP) conditional compilation macros are used to associate different lines of code to different features. 
The code snippets within \ifdef-\cppendif~blocks are only present when the \ifdef~expression evaluates to true. By setting/unsetting different macros, those expressions can evaluate to \true~or \false, and thus at preprocessing time we can vary the source code that actually gets compiled. When each of those macros corresponds to a program feature, the CPP becomes a tool for enabling/disabling individual features. In the example of Fig.~\ref{fig:annotative}, the features are $\FA$ and $\FB$, and the four product variants obtained by different feature configurations are shown on the right.

\newcommand{\TC}{\texttt{tokenCount}}

Assume we have a simple software analysis tool \TC~that, given a lexically tokenized program as input, returns the number of tokens.  
This is a straightforward counting algorithm in the case of single products. However, given an SPL, the analysis algorithm needs to simultaneously count the tokens belonging to different product variants. For the SPL in Fig.~\ref{fig:annotative}, the number of tokens is 13, 15, 17 or 18, depending on which product variant we are applying \TC~to. 

Our goal is to automatically rewrite \TC~into $\lift{\TC}$, which returns a set of all four counts, each annotated with its corresponding feature configuration. Furthermore, we require that $\lift{\TC}$~does an \emph{efficient} computation of the counts by exploiting the shared parts of the SPL. For example, all four products of the SPL in Fig.~\ref{fig:annotative} contain  "\texttt{int foo}" as a prefix, and thus its tokens should be counted only once and used as part of producing all four counts.

We present two approaches to lifting, \emph{shallow} and \emph{deep}:

\term{Shallow lifting} takes analysis \ap~as a black-box, and only wraps it in \lap~which takes variability-aware parameters and passes their combinations down to \ap. This approach is light-weight and in many cases more efficient than brute-force analysis of each product variant individually. However, it does not leverage the opportunity to share common intermediate values across different input combinations throughout the analysis.
\term{Deep lifting}, on the other hand, is a program rewriting technique that syntactically transforms a program into a semantically equivalent variability-aware one. We present rewrite rules for different syntactic constructs of the input language, and show that those rules together can be used to compositionally generate a semantically equivalent  variability-aware program.

We define the rewriting algorithm for programs written in an extended version of Programming Computable Functions (PCF)~\cite{Plotkin:1977}, which is essentially Typed Lambda Calculus (TLC) plus a fixed point operator. PCF is a concise, Turing-complete, functional programming formalism, and thus it is as computationally expressive as any high-level programming language. Several recent program rewriting and program analysis projects, for example~\cite{Kavvos:2020, Keidel:2019, Schopp:2017, Aguirre:2017}, use PCF (or variants of it) as a source language because of its expressiveness and its relatively small set of syntactic constructs.

\vskip 0.1in

\noindent
{\bf Contributions and Organization.}
In this paper, we make the following contributions:
\begin{enumerate} 
\item We define the correctness criteria for variability-aware functional programs.
\item We present a light-weight technique for lifting functional programs (shallow lifting).
\item We present a rewriting technique of programs written in an expanded variant of PCF, into semantically equivalent variability-aware programs (deep lifting). 
\item We provide sketches of correctness proofs of lifted function application.
\item We outline the design of a Haskell implementation of both shallow and deep lifting.
\item We evaluate shallow and deep lifting against brute-force SPL analysis, using four program analyses applied to the BusyBox product line of command-line tools.
\end{enumerate}

The rest of this paper is organized as follows.  Sec.~\ref{sec:background} provides background
and definitions.
Our shallow lifting framework is described in Sec.~\ref{sec:lifting}, followed by the deep lifting framework in Sec.~\ref{sec:deep}. Correctness is then covered in Sec.~\ref{sec:correctness}. Sec.~\ref{sec:impl} outlines the implementation of both frameworks, and evaluation results are presented and discussed in Sec.~\ref{sec:evaluation}. We provide some discussion points in Sec.~\ref{sec:discussion}, compare our
approach to related work in Sec.~\ref{sec:related}, and conclude in Sec.~\ref{sec:conclusion}.

\newcommand{\pcf}{PCF}
\newcommand{\pcfp}{PCF+}
\newcommand{\val}{\texttt{v}}
\newcommand{\T}{T}
\newcommand{\cntxt}{\Gamma}
\newcommand{\eval}{\downarrow}
\newcommand{\Int}{\texttt{Int}}
\newcommand{\codelit}[1] {\texttt{#1}}
\newcommand{\codevar}[1] {\texttt{#1}}

\newcommand{\tnat}{\kw{nat}}
\newcommand{\tbool}{\kw{bool}}
\newcommand{\tlist}{[\T]}
\newcommand{\tpair}{(\T, \T)}
\newcommand{\binop}{binop}
\newcommand{\kwiszero}{\kw{iszero}}
\newcommand{\kwfix}{\kw{fix}}
\newcommand{\kwif}{\kw{if}}
\newcommand{\kwthen}{\kw{then}}
\newcommand{\kwelse}{\kw{else}}
\newcommand{\kwcase}{\kw{case}}
\newcommand{\kwof}{\kw{of}}
\newcommand{\kwarrow}{\kw{\to}}
\newcommand{\kwnil}{\kw{nil}}
\newcommand{\kwcons}{\kw{cons}}
\newcommand{\kwisnil}{\kw{isnil}}
\newcommand{\kwhead}{\kw{head}}
\newcommand{\kwtail}{\kw{tail}}
\newcommand{\kwfst}{\kw{fst}}
\newcommand{\kwsnd}{\kw{snd}}
\newcommand{\numeral}{\underline{n}}
\newcommand{\kwtrue}{\kw{true}}
\newcommand{\kwfalse}{\kw{false}}

\newcommand{\alt}{~|~}

\section{Background}
\label{sec:background}
We start by summarizing the basic Software Product Line and \pcf~concepts used in this paper.

\subsection{Software Product Lines}
Different variants of an SPL have different \term{features}. A \tm{Feature} is an externally visible attribute of a software system. Examples include pieces of functionality, support for peripheral devices, and performance optimizations.
We follow the \term{annotative} product line approach~\cite{Czarnecki:2005,Kastner:2008,Rubin:2012}, formally defined below.

\BD[SPL]
An SPL $\spl$ is a tuple $(\featset, \featmodel, \domainmodel, \pcmap)$ where:
(1) $\featset$ is the set of features s.t.  an individual product can be derived from $\spl$ via a \term{feature configuration} $\config \subseteq \featset$.
(2) $\featmodel \in \Prop(\featset)$ is a propositional formula over $\featset$ defining the valid set of feature configurations. $\featmodel$ is called a \term{Feature Model (FM)}. 
(3) $\domainmodel$ is a set of program elements, called the \term{domain model}. The whole set of program elements is sometimes referred to as the \term{150\% representation}. 
(4) $\pcmap:\domainmodel \to \Prop(\featset)$ is a total function mapping each program element to a proposition (\term{feature expression}) defined over the set of features $F$. $\pcmap(e)$ is called the \term{Presence Condition (PC)} of element $e$, i.e., the set of product configurations in which $e$ is present.
\ED


For the example in Fig.~\ref{fig:annotative}, the feature set of the product line contains two features: $\FA$ and $\FB$.
In this example, domain model elements are the individual syntactic tokens of the program. \ifdef~CPP directives are used to annotate code snippets with feature names. For example, we have \emph{four} variants of function \code{foo}. Two variants of \code{foo} (with feature $\FA$) have one parameter each (lines 4-9). The other two variants (where feature $\FA$ is absent) have two parameters each (lines 12-19).

In this example all feature combinations of $\FA$ and $\FB$ are allowed, so the feature model $\featmodel$ is equal to the proposition \TT~(i.e., no combinations are invalid). However, more realistic SPLs usually allow only a proper subset of feature combinations.

The implementation of each of the four variants of \code{foo} depends on feature $\FB$. Nesting \ifdef~directives semantically evaluates to the conjunction of individual features. Syntactically, tokens on line 6 have presence condition $(\FA \wedge \FB)$, while those on line 8 have $(\FA \wedge \neg \FB)$.

\BD[\term{Feature Configuration}] \label{def:featconfig}
A valid feature configuration $\config$ of
a product line $\spl$ is a subset of its features that satisfies $\featmodel$, i.e., $\featmodel$ evaluates to \TT~when each variable $f$ of $\featmodel$ is substituted by \TT~when $f \in \config$ and by \FF~otherwise. The set of all valid configurations in $\L$ is denoted by $\Conf(\spl)$.
\ED
In our example,
 $\Conf(\spl) = \{ \{\}, \{\FA\}, \{\FB\}, \{\FA, \FB\}\}$.

\BD[\term{Product Derivation}]
A product $M$ is \emph{derived from} the product line $\L$ under the feature configuration $\rho$
if $M$ contains those and only those elements from the domain model whose presence conditions
are satisfied by $\rho$. We denote $M$ as $\indexProd{\SPL}{\config}$.
\ED
For example, the right-hand side of Fig.~\ref{fig:annotative} has the four different products that can be derived from the the SPL on the left-hand side, each with a configuration from $\Conf(\spl)$.

\begin{figure}[t]
	\begin{mbox}{}	
		\vspace{-0.1in}
			\begin{lstlisting}[language=c,numbers=left,basicstyle=\small,xleftmargin=10pt,morekeywords=defined]
int foo(int x, int y) {
#if defined(A) && defined(B) && defined(C) // A /\ B /\ C
	return x + y;
#else              // !(A /\ B /\ C)
	return x - y;		
#endif
}
			\end{lstlisting}
	\end{mbox}
	\caption{An example of a C-language source file with three features, but only two effective combinations.}
	\label{fig:effComb}
\end{figure}

\BD[\term{Effective Combinations}]
\label{def:effComb}
The number of effective combinations of a product line $\L$ is the number of lexically unique products that can be derived $\L$ under all feature combinations.
\ED

The number of products that can be derived from a product line is exponential in the number of product line features in the worst case. However, in many cases not all feature combinations are explicitly differentiated in product line artifacts. For example, the product line in Fig.~\ref{fig:effComb} has three features, but the presence conditions (lines 2 and 4) only differentiate two sets of combinations. The presence condition $\FA \wedge \FB \wedge \FC$ on line 2 denotes the combination \{A,B,C\}, while the presence condition $\neg(\FA \wedge \FB \wedge \FC)$ on line 4 denotes the set of the remaining seven combinations, all of which are lexically identical. Thus, the number of effective combinations in this example is two.

\subsection{\pcf~and \pcfp}
\label{sec:background_PCF}

\begin{figure}[t]

\begin{subfigure}[]{\textwidth}
\begin{tabular}{ r c l r}
$T$      & ::= & $\tnat \alt \tbool \alt \tpair \alt \tlist \alt T \to T $		&	\tab (type constructors)	\\
$\Gamma$ & ::= & $\emptyset \alt \Gamma, x:T$									&	\tab (typing contexts)		\\
$\binop$ & ::= & $+ \alt - \alt * \alt /$										&   \tab (binary operators) \\
$t$      & ::= & $x \alt \lambda v:T .t \alt \kwfix~t \alt t~t$					&	\tab (lambda terms)		\\
		 &\alt & $\numeral \alt \kwtrue \alt \kwfalse $							& 	\tab (numerals and boolean constants) \\
		 &\alt & $ t~\binop~t \alt \kwiszero~t $ 								&	\tab (mathematical expressions) \\
		 &\alt & $\kwif~t~\kwthen~t~\kwelse~t$									&	\tab (conditional)	\\
		 &\alt & $\kwcase~t~\kwof~p~\kwarrow~t, ..., ~p~\kwarrow~t$					&	\tab (pattern matching) \\  
		 &\alt & $(t, t) \alt \kwfst(t) \alt \kwsnd(t)$							&	\tab (pair expressions)	\\
		 &\alt & $\kwnil \alt \kwcons~t~t \alt \kwisnil~t \alt \kwhead~t \alt \kwtail~t$ &	\tab (list expressions) \\
$p$		 & ::= & $v \alt \numeral \alt \kwtrue \alt \kwfalse \alt (p, p) \alt \kwcons~p~p$ & \tab (patterns) \\       
\end{tabular}

\caption{Syntax}
\label{fig:syntax}
\end{subfigure}

\begin{subfigure}[c]{\textwidth}

\centering
\vskip 1.5em

\begin{tabular}{l c r}
	\AxiomC{}
	\RightLabel{T-numeral}
	\UnaryInfC{$\cntxt \vdash \numeral:\tnat$}
	\DisplayProof
	
	&
	
	\AxiomC{}
	\RightLabel{T-true}
	\UnaryInfC{$\cntxt \vdash \kwtrue:\tbool$}
	\DisplayProof
	
	&
	
	\AxiomC{}
	\RightLabel{T-false}
	\UnaryInfC{$\cntxt \vdash \kwfalse:\tbool$}
	\DisplayProof
\end{tabular}

\vskip 1.5em

\begin{tabular}{l c r}
	\AxiomC{$x:T \in \cntxt$}
	\RightLabel{T-var}
	\UnaryInfC{$\cntxt \vdash x:T$}
	\DisplayProof
	
	&

	\AxiomC{$\cntxt, x:T_1 \vdash t_2:T_2$}
	\RightLabel{T-abs}
	\UnaryInfC{$\cntxt \vdash \lambda x:T_1.t_2 : T_1 \to T_2$}
	\DisplayProof

	& 
	
	\AxiomC{$\cntxt \vdash t_1:T_1 \to T_2$}
	\AxiomC{$\cntxt \vdash {t_2}:{T_1}$}
	\RightLabel{T-app}
	\BinaryInfC{$\cntxt \vdash {t_1} \: {t_2} : {T_2}$}
	\DisplayProof
\end{tabular}

\vskip 1.5em

\begin{tabular}{l r}
	\AxiomC{$\cntxt \vdash t:T \to T$}
	\RightLabel{T-fix}
	\UnaryInfC{$\cntxt \vdash \kwfix~t:T$}
	\DisplayProof
	
	&

	\AxiomC{$\cntxt \vdash t_1:\tnat$}
	\AxiomC{$\cntxt \vdash t_2:\tnat$}
	\RightLabel{T-binop}
	\BinaryInfC{$\cntxt \vdash t_1~\binop~t_2:\tnat$}
	\DisplayProof
\end{tabular}

\vskip 1.5em

\begin{tabular} {l r}
	\AxiomC{$\cntxt \vdash t:\tnat$}
	\RightLabel{T-iszero}
	\UnaryInfC{$\cntxt \vdash \kwiszero~t:\tbool$}
	\DisplayProof
	
	&
	
	\AxiomC{$\cntxt \vdash t_1:\tbool$}
	\AxiomC{$\cntxt \vdash t_2:T$}
	\AxiomC{$\cntxt \vdash t_3:T$}
	\RightLabel{T-cond}
	\TrinaryInfC{$\cntxt \vdash \kwif~t_1~\kwthen~t_2~\kwelse~t_3:T$}
	\DisplayProof

\end{tabular} 

\vskip 1.5em

\begin{tabular} {l r}
	\AxiomC{$\cntxt \vdash t:T_1$}
	\AxiomC{$\cntxt \vdash t_0 : T_2~...~\cntxt \vdash t_n:T_2$}
	\RightLabel{T-case}
	\BinaryInfC{$\cntxt \vdash \kwcase~t~\kwof~p_0 \to t_0~,...,~p_n \to t_n : T_2$}
	\DisplayProof
	
	&
	
	\AxiomC{$\cntxt \vdash t_1:T_1$}
	\AxiomC{$\cntxt \vdash t_2:T_2$}
	\RightLabel{T-pair}
	\BinaryInfC{$\cntxt \vdash (t_1, t_2) : (T_1, T_2)$}
	\DisplayProof

\end{tabular} 

\vskip 1.5em

\begin{tabular} {l r}
		
	\AxiomC{$\cntxt \vdash (t_1, t_2) : (T_1, T_2)$}
	\RightLabel{T-fst}
	\UnaryInfC{$\cntxt \vdash \kwfst(t_1, t_2) : T_1$}
	\DisplayProof
	
	&
	
	\AxiomC{$\cntxt \vdash (t_1, t_2) : (T_1, T_2)$}
	\RightLabel{T-snd}
	\UnaryInfC{$\cntxt \vdash \kwsnd(t_1, t_2) : T_2$}
	\DisplayProof
	
\end{tabular} 

\vskip 1.5em

\begin{tabular}{l c r}
	\AxiomC{}
	\RightLabel{T-nil}
	\UnaryInfC{$\cntxt \vdash \kwnil:\tlist$}
	\DisplayProof

	&
	
	\AxiomC{$\cntxt \vdash t_1:T$}
	\AxiomC{$\cntxt \vdash t_2:\tlist$}
	\RightLabel{T-cons}
	\BinaryInfC{$\cntxt \vdash \kwcons~t_1~t_2:\tlist$}
	\DisplayProof	

	&
	
	\AxiomC{$\cntxt \vdash t:\tlist$}
	\RightLabel{T-isnil}
	\UnaryInfC{$\cntxt \vdash \kwisnil~t:\tbool$}
	\DisplayProof
\end{tabular}

\vskip 1.5em

\begin{tabular}{l r}
	\AxiomC{$\cntxt \vdash t:\tlist$}
	\RightLabel{T-head}
	\UnaryInfC{$\cntxt \vdash \kwhead~t:T$}
	\DisplayProof

	&
	
	\AxiomC{$\cntxt \vdash t:\tlist$}
	\RightLabel{T-tail}
	\UnaryInfC{$\cntxt \vdash \kwtail~t:\tlist$}
	\DisplayProof
\end{tabular}

\caption{Typing rules.}
\label{fig:typing}
\end{subfigure}

\caption{\pcfp~syntax and typing rules.}
\label{fig:PCF}
\vspace{-0.1in}
\end{figure}

\begin{figure}[t]
	\vskip 1.5em
	
	\centering
	\begin{tabular}{l r}
		\AxiomC{}
		\RightLabel{E-numeral}
		\UnaryInfC{$\underline{n} \eval n$}	
		\DisplayProof
		
		&
		
		\AxiomC{$M \eval m$}
		\AxiomC{$N \eval n$}
		\RightLabel{E-binop}
		\BinaryInfC{$M~\binop~N \eval m~\semantics{binop}~n$}
		\DisplayProof 
		
	\end{tabular}

	\vskip 1.5em
	
	\begin{tabular}{c}
		\AxiomC{$M \eval \lambda x . P$}
		\AxiomC{$P[N/x] \eval V$}
		\RightLabel{E-app}
		\BinaryInfC{$M N \eval V$}	
		\DisplayProof

	\end{tabular}
	
	\vskip 1.5em
	
	\begin{tabular}{l r}
		\AxiomC{$t_1 \eval t_2$}
		\RightLabel{E-fix}
		\UnaryInfC{$\kwfix~t_1 \eval \kwfix~t_2$}	
		\DisplayProof 
		
		&
		
		\AxiomC{}
		\RightLabel{E-fixBeta}
		\UnaryInfC{$\kwfix (\lambda x:T.t) \eval t[\kwfix (\lambda x:T.t)/x]$}	
		\DisplayProof 
		
	\end{tabular}
	
	\vskip 1.5em
	
	\begin{tabular} {l r}
		\AxiomC{$t \eval 0$}
		\RightLabel{E-iszero1}
		\UnaryInfC{$\kwiszero~t \eval \kwtrue$}	
		\DisplayProof 
		
		&
		
		\AxiomC{$t \eval v$}
		\AxiomC{$v \neq 0$}
		\RightLabel{E-iszero2}
		\BinaryInfC{$\kwiszero~t \eval \kwfalse$}	
		\DisplayProof 
		
	\end{tabular}

	\vskip 1.5em
	
	\begin{tabular}{l r}
		\AxiomC{$B \eval \kwtrue$}
		\AxiomC{$M \eval v$}
		\RightLabel{E-cond1}
		\BinaryInfC{$\kwif~B~\kwthen~M~\kwelse~N \eval v$}	
		\DisplayProof  
		
		&
		
		\AxiomC{$B \eval \kwfalse$}
		\AxiomC{$N \eval v$}
		\RightLabel{E-cond2}
		\BinaryInfC{$\kwif~B~\kwthen~M~\kwelse~N \eval v$}	
		\DisplayProof 
		
	\end{tabular}
	
	\vskip 1.5em
	
	\begin{tabular}{c}
		\AxiomC{$t \eval u$}
		\AxiomC{$match(p_i, u) \land \forall_{j < i}: \neg match(p_j, u)$}
		\AxiomC{$t_i[v_0/x_0,...,v_k/x_k] \eval w$}
		\RightLabel{E-case}
		\TrinaryInfC{$\kwcase~t~\kwof~p_0 \to t_0, ..., p_n \to t_n \eval w$}	
		\DisplayProof
		
	\end{tabular}

	\vskip 1.5em

	\begin{tabular}{l r}
		\AxiomC{$M \eval \kwnil$}
		\RightLabel{E-isnil1}
		\UnaryInfC{$\kwisnil~M \eval \kwtrue$}	
		\DisplayProof  
		
		&
		
		\AxiomC{$M \eval \kwcons~N~P$}
		\RightLabel{E-isnil2}
		\UnaryInfC{$\kwisnil~M \eval \kwfalse$}	
		\DisplayProof 
		
	\end{tabular}
	
	\vskip 1.5em
	
	\begin{tabular}{l c r} 
		
		\AxiomC{$M \eval v$}
		\RightLabel{E-fst}
		\UnaryInfC{$\kwfst~(M, N) \eval v$}	
		\DisplayProof
		
		&
		
		\AxiomC{$N \eval v$}
		\RightLabel{E-snd}
		\UnaryInfC{$\kwfst~(M, N) \eval v$}	
		\DisplayProof
		
		&
		
		\AxiomC{$M \eval v$}
		\RightLabel{E-head}
		\UnaryInfC{$\kwhead~(\kwcons~M~N) \eval v$}	
		\DisplayProof
	\end{tabular}

	\vskip 1.5em
	
	\begin{tabular}{l c r}
		\AxiomC{}
		\RightLabel{E-tail1}
		\UnaryInfC{$\kwtail~\kwnil \eval \kwnil$}	
		\DisplayProof 
		
		&
		
		\AxiomC{$N \eval v$}
		\RightLabel{E-tail2}
		\UnaryInfC{$\kwtail~(\kwcons~M~N) \eval v$}	
		\DisplayProof
	\end{tabular}
	
	\caption{Call-By-Name (CBN) Operational semantics of \pcfp.}
	\label{fig:semantics}
	\vspace{-0.1in}
\end{figure}

\term{Programming Computable Functions (\pcf)}~\cite{Plotkin:1977} is a simple language based on Typed Lambda Calculus, plus a fixed point operator for general recursion. \pcf~is Turing-complete, but still minimal in terms of constructs and data types. We begin by extending \pcf~ to add polymorphic pairs and lists (to demonstrate lifting of product and sum types), and pattern matching using $\kwcase$ expressions (along the lines of ~\cite{Mitchell:1996, Pierce:2002}).  We refer to the resulting
language as \pcfp.  Fig.~\ref{fig:PCF} summarizes the \pcfp~syntax and typing rules. \pcfp~Call By Name (CBN) semantics is summarized in Fig.~\ref{fig:semantics}. 

{\bf Syntax.} A syntactic term of \pcfp~(Fig.~\ref{fig:syntax}) is either a variable, a lambda abstraction, a recursive function definition (using a fixed point operator), lambda application, a numeral (underlined to distinguish it from its semantic natural number), a mathematical expression (using addition, subtraction, multiplication and division operators), a Boolean constant ($\kwtrue$ or $\kwfalse$), a conditional expression, a $\kwcase$ expression (pattern matching), a constructor expression of a pair, or a constructor expression of a list. The language includes primitives for pair and list manipulation ($\kwfst$, $\kwsnd$, $\kwisnil$, $\kwhead$, and $\kwtail$). Each well-typed syntactic term has a unique type that is either $\tnat$, $\tbool$, $(\T_1, \T_2)$, $\tlist$, or a unary function type $(\T_1 \to \T_2)$ from type $\T_1$ to type $\T_2$. The language supports functions of higher arities using currying.

{\bf Typing rules.} Given a global typing context $\cntxt$, each well-typed term is assigned a type according to the typing rules of Fig.~\ref{fig:typing}. All numeral constants have type $\tnat$, and Boolean constants $\kwtrue$~and $\kwfalse$ have type $\tbool$. Variable types are stored in the typing context $\cntxt$. A lambda abstraction from type $\T_1$ to type $\T_2$ has type $\T_1 \to \T_2$. A lambda application term $t_1 t_2$ has type $\T_2$ if $t_1$ has type $\T_1 \to \T_2$ and $t_2$ has type $\T_1$. A fixed point over a function of type $\T \to \T$ has type $\T$ (the type system does not designate a type for diverging functions). 

Applying an infix binary operator to two terms of type $\tnat$~is a term of type $\tnat$~(again, exceptions like division by zero are not designated a type). Applying the $\kwiszero$ unary operator to a term of type $\tnat$ results in a term of type $\tbool$. An $\kwif$-$\kwthen$-$\kwelse$~term is of type $\type{T}$ if both the $\kwthen$-term and $\kwelse$-term are of type $\T$, and the condition is of type $\tbool$. A $\kwcase$ expression has type $T_2$ if 
each of its alternatives is an expression of type $\T_2$. 

The pair type has one constructor, taking two arguments of types $\T_1$ and $\T_2$, respectively, and constructing a value of type $(\T_1, \T_2)$. A list over type $\T$ has two constructors: $\kwnil$ constructing an empty list, and $\kwcons$ constructing a list from a value of type $\T$ and a $\tlist$. The $\kwisnil$ unary operator takes a $\tlist$ and returns a $\tbool$. Operators $\kwhead$~and $\kwtail$ each takes a $\tlist$ and returns a $\T$ and a $\tlist$, respectively.

{\bf Operational semantics.} Fig.~\ref{fig:semantics} outlines the \term{Call-By-Name (CBN)} operational semantics of PCF. Constants ($\kwtrue$, $\kwfalse$, and $\kwnil$) are irreducible normal forms. A numeral $\numeral$ reduces to its corresponding natural number. A binary mathematical operator symbol $\binop$ reduces to its corresponding mathematical operator $\semantics{\binop}$, so when it is used as an infix operator on two terms reducing to natural numbers $m$ and $n$, it reduces to the result of the mathematical expression $m~\semantics{binop}~n$.

Function application substitutes the argument in place of the parameter in the function body. A fixed point operator applies its argument to itself (allowing for the definition of recursive functions). The $\kwiszero$ operator returns $\kwtrue$ if its argument reduces to zero, and returns $\kwfalse$ otherwise. An $\kwif$-$\kwthen$-$\kwelse$~term reduces to what the $\kwthen$-term reduces to if the condition reduces to $\kwtrue$, and reduces to what the $\kwelse$-term reduces to otherwise.

A $\kwcase$ expression evaluates its argument $t$ to value $u$, and tries to match the structure of $u$ against each of the patterns one-by-one (in order). The alternative expression $t_i$ of the first matching pattern $p_i$ is then reduced to value $w$, substituting each of the pattern variables $x$ with its corresponding matching value $v$ from $u$. The value $w$ is what the overall $\kwcase$ expression reduces to.

Expressions $(\kwfst~M)$ and $(\kwsnd~M)$ reduce to values $x$ and $y$, respectively, if $M$ reduces to $(x,y)$. Operator $\kwisnil$  applied to a term $M$ reduces to $\kwtrue$ if $M$ reduces to $\kwnil$, and reduces to $\kwfalse$ otherwise. Operator $\kwhead$  applied to a term $(\kwcons~M~N)$ reduces to $v$ if $M$ reduces to $v$. Operator $\kwtail$ reduces to $\kwnil$ if applied to $\kwnil$, and  to $v$ if applied to $(\kwcons~M~N)$ and $N$ reduces to $v$.

\pcf~is Turing-complete~\cite{Plotkin:1977}, allowing for the definition of partial functions, so we assume undefined behavior (e.g., division by zero, head of an empty list) to be semantically equivalent to diverging expressions.

%

\newcommand{\PCType}{\texttt{PC}}
\newcommand{\pcval}{\texttt{pc}}
\newcommand{\Tlifted}{\lift{\T}}
\newcommand{\lit}[1]{\underline{#1}}

\section{Shallow Lifting}
\label{sec:lifting}

\begin{figure}[t]
\scalebox{0.8}{}
		\centering
		\begin{tabular}{l l c l}
		
		1 & tokenCount & = & $\kwfix~\lambda x:[\tnat]~.$ \\
		2 & 		   &   & \tab $\kwcase~x~\kwof$ \\
		3 & 		   &   & \tab \tab $\kwnil~\to~\underline{0}$ \\
		4 & 		   &   & \tab \tab $\kwcons~h~t~\to$ \\
		5 & 		   &   & \tab \tab \tab $\kwif~\kwiszero~h$ \\
		5 & 		   &   & \tab \tab \tab $\kwthen$ tokenCount $t$ \\
		6 & 		   &   & \tab \tab \tab $\kwelse~\underline{1} +$ tokenCount $t$
		\end{tabular}
		
		
		
	
	\caption{\TC~analysis.}
	\label{fig:tokenCount}
\end{figure}

%
%
%
%
%
%



Consider the token counting example from Fig.~\ref{fig:annotative}. Assume each token has a unique non-zero natural number identifier, where identifier \underline{0} corresponds to a special empty token. Fig.~\ref{fig:tokenCount} shows a \pcfp~implementation of the \TC~analysis introduced in Sec.~\ref{sec:intro}. This analysis cannot take an SPL (like the one in Fig.~\ref{fig:annotative}) as input because it cannot handle variability. A variability-aware version of \TC~(let us call it $\lift{\TC}$) needs to take a variability-aware list as input, return a variability-aware result (token counts for each of the program variants), and in order to compute the correct results, it has to systematically track intermediate values and associate each of them to its corresponding variant. This all has to be done while maintaining the semantics of the original \TC~analysis.

Our goal is to automatically lift an arbitrary \pcfp~program $\prog$ to its semantically equivalent lifted program $\lift{\prog}$. In this section, we achieve that goal by taking $\prog$ as a black-box, and wrapping it in $\lift{\prog}$. We call this approach \term{shallow lifting}. 

\subsection{Variability-Aware Values}

In \pcfp, a value \val~of type \T~is a singleton belonging to that type. For example, \underline{7} is a value of type $\tnat$. In variability-aware domains, on the other hand, a variable of type \T~can simultaneously have different values of type \T, each in a different set of products identified by a \term{Presence Condition (PC)}.

For example, the SPL in Fig.~\ref{fig:annotative} has two features, $\FA$ and $\FB$, and PCs are Boolean expressions over those two features. The product space is partitioned into four subsets labeled by the PCs $(\FA \wedge \FB)$,  $(\FA \wedge \neg \FB)$, $(\neg \FA \wedge \FB)$, and $(\neg \FA \wedge \neg \FB)$. 
The result of our token count analysis differs from one set of products to another. For the example in Fig.~\ref{fig:annotative}, the number of tokens is \lit{13} in $(\FA \wedge \FB)$, \lit{15} in $(\FA \wedge \neg \FB)$, \lit{18} in $(\neg \FA \wedge \FB)$, and \lit{17} in $(\neg \FA \wedge \neg \FB)$.
That variability-aware value is the set of $\tnat$-\PCType~pairs $\{(\lit{13}, \FA \wedge \FB), (\lit{15}, \FA \wedge \neg \FB), (\lit{18}, (\neg \FA \wedge \FB)), (\lit{17}, \neg \FA \wedge \neg \FB)\}$. Each pair maps a $\tnat$~value (we call that an \term{atomic value}) to a set of products denoted by a presence condition.

We assume that the \PCType~type encapsulates presence conditions over the feature space. A variability-aware value over a type \T~is a set of pairs (\val, \pcval), where \val~is of type \T~and \pcval~is of type \PCType, and the following \term{disjointness} and \term{full coverage} invariants hold.

\newtheorem*{disjInv}{Disjointness Invariant}{\bfseries}{\itshape}
\begin{disjInv}
	Given a variability-aware value $\lift{\val}$:
	\begin{equation}
	\lift{\val} = \{(\val_1, \pc{1}), ..., (\val_n, \pc{n})\}, \forall i \neq j: {\unsat (\pc{i} \wedge \pc{j})}
	\end{equation}
\end{disjInv}

The disjointness invariant intuitively states that a variability-aware value can have at most one atom value in any product configuration. The conjunction of two presence conditions is unsatisfiable if and only if they represent non-overlapping sets of products. Without this invariant, the semantics of variability-aware values would be non-deterministic.
This invariant is a precondition on program inputs, and in Sec.~\ref{sec:correctness} we show how its validity is maintained by the lifting transformations.

The second invariant states that a variability-aware value $\lift{\val}$ covers the full product space, not leaving out any valid feature combination.

\newtheorem*{covInv}{Full Coverage Invariant}{\bfseries}{\itshape}
\begin{covInv}
	\begin{equation}
	\lift{\val} = \{(\val_1, \pc{1}), ..., (\val_n, \pc{n})\}, \bigvee_{i} \pc{i} = \featmodel
	\end{equation}
\end{covInv}

The two invariants together ensure that, semantically, a variability-aware value $\lift{\val}$~of type \T~is a total mapping from the set of valid product configurations to the underlying data type \T:
\[ \typerel{\semantics{\typerel{\lift{\val}}{\lift{\T}}}} {\featmodel \to \T} \]

\BD[\term{Lifted Value Indexing}]
Given a lifted value $\lift{v}$ and a configuration $\config$, $\indexProd{\lift{v}}{\config}$ is the atomic value of $\lift{v}$ whose presence conditions is satisfied by $\config$.
\ED

\begin{example}[Lifted Variables]
	Given features $\FA$ and $\FB$, variables \codevar{x}, \codevar{y}, and \codevar{z} of type $\lift{\tnat}$ can have the following values: 
	
	\codevar{x} = \{(\lit{1},$\FA$),(\lit{2}, $\neg \FA \wedge \FB$),(\lit{0}, $\neg \FA \wedge \neg \FB$)\}
	
	\codevar{y} = \{(\lit{5}, $\FA \wedge \neg \FB$), (\lit{4},$\FB$), (\lit{3}, $\neg \FA \wedge \neg \FB$)\}
	
	\codevar{z} = \{(\lit{19}, $\TT$)\}
	
	\label{ex:var}
	
	Variable \codevar{x} has the value \lit{1} when feature $\FA$ is present. If $\FA$ is absent, \codevar{x} is \lit{2} if $\FA$ is present, and \lit{0} if $\FB$ is absent. Variable \codevar{y} has the value \lit{4} if $\FB$ is present, \lit{5} if $\FB$ is absent and $\FA$ is present, and \lit{3} if both features are absent. Variable \codevar{z} has the value 19 in all products regardless of which features are present or absent. 
	
	Examples of applying the value indexing operator are $\indexProd{\codevar{x}}{\{A\}} = \lit{1}$ ($\FA$ is present and $\FB$ is absent), $\indexProd{\codevar{y}}{\{A,B\}} = \lit{4}$ (both $\FA$ and $\FB$ are present), and $\indexProd{\codevar{z}}{\{\}} = \lit{19}$ (none of the two features is present).
	
\end{example}

\subsection{Variability-Aware Types}
 
Given a type \T, its corresponding variability-aware type $\lift{\T}$ is the type of sets of \T-\PCType~pairs satisfying both the disjointness and completeness invariants:

\begin{prooftree}
	\AxiomC{$\typerel{\val_1, \ldots , \val_n}{\T}, \pc{1}, \ldots , \pc{n} : \PCType$}
	\AxiomC{$\forall i \neq j \cdot \unsat (pc_i \wedge pc_j)$}
	\AxiomC{$\bigvee_{i} {pc_i} = \featmodel $}
	\RightLabel{lifted-type}
	\TrinaryInfC{$\typerel { \{(\val_1,\pc{1}), \ldots , (\val_n, \pc{n})\} } { \Tlifted } $}	
\end{prooftree}

Note that we are only restricting input programs to be in \pcfp, but we can rewrite them into a language with a richer type system with polymorphic types and type classes (e.g., Haskell). In Sec.~\ref{sec:impl}, we show how lifted types can be implemented as a Haskell polymorphic type.

\newcommand{\f}{\codelit{f}}
\newcommand{\atob}{(\codelit{a} \to \codelit{b})}
\newcommand{\apply}{\codelit{apply}}

\subsection{Lifting Function Application}
In \pcfp~(and functional programming in general), functions are values. Following our treatment of lifted values, a function \f~of type $\atob$~is lifted to $\lift{\codelit{f}}$~of type $\lift{\atob}$. 
However, following the \pcfp~type system rules, $\lift{\codelit{f}}$~should be of type $\lift{\codelit{a}} \to \lift{\codelit{b}}$ if we are to apply it to values of type $\lift{\codelit{a}}$ and get values of type $\lift{\codelit{b}}$ back. As a result, we need to introduce our own lifted function application mechanism on top of \pcfp~function application.

Starting with the simplest case, where we have a lifted unary function of type $\atob$ and an argument of type $\lift{\codelit{a}}$, we need to provide a way to apply that function to its argument resulting in a $\lift{\codelit{b}}$ value.
Recall that a lifted value of type $\lift{\codelit{t}}$ is just a set of values of type \codelit{t}, each paired with a PC (with the whole set satisfying the disjointness and full coverage invariants). To apply a collection of functions to a collection of values, we only need to apply each of the functions to each of the values. Since the result has to be a lifted value, we also need a PC for each of the computed values. Intuitively, since the result exists only if its generating function and argument exist, then its PC is the conjunction of the PCs of the function and the argument. If the conjunction of the PCs is unsatisfiable (i.e., a logical contradiction), it means that the set of products where the function is defined and that where the atomic value is defined do not intersect. Consequently, the result can be safely filtered out. 

The \apply~operator does exactly that. It generates the cross product of functions and arguments from \f~and \codelit{x}. For each element in the cross product (a pair of pairs), it conjoins the individual PCs, and filters the element out if the conjunction is not satisfiable. The semantics of \apply~is summarized by the following inference rule:

\begin{prooftree}
\AxiomC{$\typerel{\f}{\lift{\atob}}$}
\AxiomC{$\typerel{\codelit{x}} {\lift{\codelit{a}}}$}
\RightLabel{$\apply$}
\BinaryInfC{$(\typerel{\apply~\f~\codelit{x}}{\lift{\codelit{b}}}) = \{(\f'(\codelit{x}'), \codelit{fpc} \wedge \codelit{xpc})~|~(\f', \codelit{fpc}) \in \f, (\codelit{x}', \codelit{xpc}) \in \codelit{x}, \sat(\codelit{fpc} \wedge \codelit{xpc})\}$}	
\end{prooftree}

\begin{example}[Unary function application]
Assume we need to apply the $\kwiszero$ function to the lifted variable \codevar{x} from the Example~\ref{ex:var}. First, we need a lifted version of $\kwiszero$. Since the behavior of $\kwiszero$ does not vary across different product configurations, we can just wrap it into a singleton lifted value with presence condition \TT: $\lift{\kwiszero} = \{(\kwiszero, \text{\TT})\}$.

Applying $\lift{\kwiszero}$ to \codevar{x} evaluates to:

\begin{tabular}{l c l l}
	$\apply~\lift{\kwiszero}~\codevar{x}$ & = & $\{$ & $(\kwiszero~\lit{1}, \TT \land \FA),$ \\
	                                      &   &      & $(\kwiszero~\lit{2}, \TT \land \neg \FA \land \FB),$ \\
	                                      &   &      & $(\kwiszero~\lit{0}, \TT \land \neg \FA \land \neg \FB)\}$ \\
	                                      & = & $\{$ & $(\kwfalse, \FA),$ \\
                                          &   &      & $(\kwfalse, \neg \FA \land \FB),$ \\
                                          &   &      & $(\kwtrue, \neg \FA \land \neg \FB)\}$ \\
\end{tabular}

\end{example}

\begin{example}[Binary function application]
Now assume that we would like to add \codevar{x} and \codevar{y} from Example~\ref{ex:var}. The first step is to get a lifted version of the \codevar{(+)} binary operator: $\lift{\codevar{(+)}} = \{(\codevar{(+)}, \TT)\}$.

We cannot directly use \apply~ because it expects only two arguments, the first of which is a lifted unary function. However, functions in \pcf~are all unary, and functions of higher arities are just Curried higher-order functions \cite{Mitchell:1996}. If we pass $\lift{\codevar{(+)}}$ and \codevar{x} to \apply, it will apply the binary lifted operator $\lift{\codevar{(+)}}$ to a single lifted argument \codevar{x}:

\[(\codevar{i}:\lift{(\tnat \to \tnat)}) = \apply~\lift{\codevar{(+)}}~\codevar{x}\]

Since multi-parameter functions are Curried unary functions, the intermediate value \codevar{i} is a lifted unary function value of type $\lift{(\tnat \to \tnat)}$, evaluating to:

\begin{tabular}{l c l l}
	$\codevar{i} = \apply~\lift{\codevar{(+)}}~\codevar{x}$ & = & $\{$ & $(\lift{\codevar{(+)}}~\lit{1}, \TT \land \FA),$ \\
	&   &      & $(\lift{\codevar{(+)}}~\lit{2}, \TT \land \neg \FA \land \FB),$ \\
	&   &      & $(\lift{\codevar{(+)}}~\lit{0}, \TT \land \neg \FA \land \neg \FB)\}$ \\
	& = & $\{$ & $((\lit{1} + ), \FA),$ \\
	&   &      & $((\lit{2} + ), \neg \FA \land \FB),$ \\
	&   &      & $((\lit{0} + ), \neg \FA \land \neg \FB)\}$ \\
\end{tabular}

Now we just need to apply \codevar{i}, which is a lifted unary function of type $\lift{(\tnat \to \tnat)}$, to \codevar{y}:

\begin{tabular}{l c l l}
	$\apply~\codevar{i}~\codevar{y}$ & = & $\{$ & $((\lit{1} + )~\lit{5}, \FA \land \FA \land \neg \FB),$ \\
	&   &      							        & $((\lit{1} + )~\lit{4}, \FA \land \FB),$ \\
	&   &      							        & $((\lit{1} + )~\lit{3}, \FA \land \neg \FA \land \neg \FB),$ \\
	&   &                                       & $((\lit{2} + )~\lit{5}, \neg \FA \land \FB \land \FA \land \neg \FB),$ \\
	&   &      							        & $((\lit{2} + )~\lit{4}, \neg \FA \land \FB \land \FB),$ \\
	&   &      							        & $((\lit{2} + )~\lit{3}, \neg \FA \land \FB \land \neg \FA \land \neg \FB),$ \\
	&   &                                       & $((\lit{0} + )~\lit{5}, \neg \FA \land \neg \FB \land \FA \land \neg \FB),$ \\
	&   &      							        & $((\lit{0} + )~\lit{4}, \neg \FA \land \neg \FB \land \FB),$ \\
	&   &      							        & $((\lit{0} + )~\lit{3}, \neg \FA \land \neg \FB \land \neg \FA \land \neg \FB),$ \\
	                                & = & $\{$  & $(\lit{6}, \FA \land \neg \FB),$ \\
	&   &      							        & $(\lit{5}, \FA \land \FB),$ \\
	&   &      							        & $(\lit{4}, \FF),$ \\
	&   &                                       & $(\lit{7}, \FF),$ \\
	&   &      							        & $(\lit{6}, \neg \FA \land \FB),$ \\
	&   &      							        & $(\lit{5}, \FF),$ \\
	&   &                                       & $(\lit{5}, \FF),$ \\
	&   &      							        & $(\lit{4}, \FF),$ \\
	&   &      							        & $(\lit{3}, \neg \FA \land \neg \FB)\}$ \\
\end{tabular}

Eliminating the pairs with \FF~(unsatisfiable) presence conditions, we are left with exactly four pairs, one for each product configuration. 

Versions of \apply~of different arities can thus be defined using straightforward Currying. For example, the implementation of \apply2 of type $\lift{(\codelit{a} \to \codelit{b} \to \codelit{c})} \to \lift{\codelit{a}} \to \lift{\codelit{b}} \to \lift{\codelit{c}}$ is
\[ \apply2~\f~\codevar{x}~\codevar{y} = \apply~(\apply~\f~\codevar{x})~\codevar{y}.\]
\end{example}

\newcommand{\ffoo}{\codevar{foo}}
\newcommand{\fbar}{\codevar{bar}}
\newcommand{\fbaz}{\codevar{baz}}
\newcommand{\x}{\codevar{x}}
\newcommand{\y}{\codevar{y}}
\newcommand{\z}{\codevar{z}}

\begin{figure}[t]
	\begin{tabular}{lc}
		\begin{tabular}{l}
			\begin{subfigure}[]{0.59\textwidth}
				
				\begin{tabular}{l c l}	
					$\ffoo$ & = & $\lambda \x:\tnat.~\lambda \y:\tnat.~\lambda \z:\tnat.$ \\
					&   & $(\fbar~\x~\y) + (\fbaz~\z)$
				\end{tabular}
				
				\caption{Function \codevar{foo}.}
				\label{fig:foo}
			\end{subfigure}
			\vspace{0.1in}
			\\
			\begin{subfigure}[]{0.59\textwidth}
				
				\codevar{x} = \{(\lit{-7},$\FA$),(\lit{3}, $\neg \FA$)\}
				
				\codevar{y} = \{(\lit{1}, $\FA \wedge \FB$), (\lit{8},$\FA \wedge \neg \FB$), (\lit{4}, $\neg \FA \wedge \FB$), (\lit{10}, $\neg \FA \wedge \neg \FB$)\}
				
				\codevar{z} = \{(\lit{5}, $\TT$)\}
				
				\caption{Arguments \codevar{x}, \codevar{y}, and \codevar{z}.}
				\label{fig:args}
			\end{subfigure}
			\vspace{0.1in}
			\\
			\begin{subfigure}[]{0.59\textwidth}
				
				\begin{tabular}{l c l}	
					$\lift{\ffoo}$  & = & \{(\ffoo, \TT)\} \\
					\codevar{result}& = & \apply3 $\lift{\ffoo}~\x~\y~\z$
				\end{tabular}
				
				\caption{Shallow-lifted $\lift{\ffoo}$ applied to $\x$, $\y$, and $\z$.}
				\label{fig:foolifted}
			\end{subfigure} 
		\end{tabular}
		&
		\begin{subfigure}[]{0.3\textwidth}
			\begin{tabular}{cccl}
				\toprule
				$\x$ & $\y$ & $\z$ & \textit{\textbf{PC}} \\
				\midrule
				-7 & 1 & 5 & $\FA \wedge \FB$ \\
				-7 & 8 & 5 & $\FA \wedge \neg \FB$ \\
				\st{-7} & \st{4} & \st{5} & \st{$\FF$} \\
				\st{-7} & \st{10} & \st{5} & \st{$\FF$} \\
				\st{3} & \st{1} & \st{5} & \st{$\FF$} \\
				\st{3} & \st{8} & \st{5} & \st{$\FF$} \\
				3 & 4 & 5 & $\neg \FA \wedge \FB$ \\
				3 & 10 & 5 & $\neg \FA \wedge \neg \FB$ \\
				\bottomrule
			\end{tabular}
			\caption{Input vectors for $\lift{\codevar{foo}}$.}
			\label{fig:inputs}	
		\end{subfigure}
	\end{tabular}
	\caption{Shallow lifting $\ffoo$ into $\lift{\ffoo}$, and applying $\lift{\ffoo}$ to $\x$, $\y$, and $\z$. Fig.(d) lists the input vectors passed to $\ffoo$, with those with unsatisfiable PCs crossed out.}
	\vspace{-0.1in}
	\label{fig:fooExample}
\end{figure}

Shallow lifting wraps a function in its lifted counterpart. While straightforward, it treats the original function as a black-box, so it does not leverage opportunities for reuse of common intermediate values within that function. For example, function $\ffoo$ in Fig.~\ref{fig:foo} takes three $\tnat$ arguments, passing the first two to function $\fbar$ and the third to function $\fbaz$. If we are to call the shallow-lifted function $\lift{\ffoo}$ (Fig.~\ref{fig:foolifted}) with arguments in Fig.~\ref{fig:args}, the original function $\ffoo$ gets called with the input vectors in Fig.~\ref{fig:inputs} (those with unsatisfiable PCs are crossed out). 

The column for argument $\z$ has the value 5 for all input vectors. As a result, $\ffoo$ will internally call $\fbaz$ four times, each with the exact same argument 5. This is an example of a case where an intermediate value should be computed only once and subsequently reused. Since shallow lifting does not introspect the structure of the function being lifted, those value sharing opportunities are wasted.

\newcommand{\rewrite}{\Rightarrow}

\section{Deep Lifting}
\label{sec:deep}

Deep lifting rewrites the body of a function, pushing lifted function application down into the function body. For example, a deep-lifted $\ffoo$ (from Fig.~\ref{fig:fooExample}) would look like this:

\begin{tabular}{l c l}	
	$\lift{\ffoo}$ & = & $\lambda \x:\lift{\tnat}.~\lambda \y:\lift{\tnat}.~\lambda \z:\lift{\tnat}.$ \\
	               &   & $\apply2~\lift{(+)}$ $(\apply2~\lift{\fbar}~\x~\y)$ $(\apply~\lift{\fbaz}~\z)$ \\
\end{tabular}

Assuming $\lift{\fbar}$ and $\lift{\fbaz}$ are shallow-lifted versions of $\fbar$ and $\fbaz$ respectively, redundant calls to them will be eliminated. For example, since $\z$ is a singleton, applying $\lift{\fbaz}$ to it would result in only one call to the underlying function $\fbaz$ instead of four redundant calls as in the case of shallow-lifted $\ffoo$.

In this section, we present the rewrite rules for the syntactic constructs of \pcfp that require recursive rewriting (primitive constructs are shallow-lifted). The $\rewrite$ symbol stands for "rewrite-into".

\subsection{Lifting Function Definitions}
A function body is a \pcfp~term. Primitive values (e.g., numerals, Boolean constants) and calls to primitive built-in functions (e.g., mathematical operators) are shallow lifted because they do not have an internal structure to push lifting into. Calls to user-defined functions (e.g., \fbar~and \fbaz~in Fig.~\ref{fig:fooExample}) can be either shallow-lifted or deep-lifted. If the called function belongs to another module that is not being deep-lifted, then the function call is  shallow-lifted using the different variants of the \apply~operator. If on the other hand the callee belongs to the same module or to another module that is being deep-lifted then the call is left as is. The deep-lifted callee expects variability-aware arguments, and the parameters passed by the deep-lifted caller are variability-aware, so built-in \pcfp~function application can be used right away.


\subsection{Lifting Conditional Expressions}
In Call-By-Name(CBN) functional programming, a conditional expression is just syntactic sugar for a ternary polymorphic function, taking a Boolean argument and two arguments of type \T and returning one of them depending on what the first argument evaluates to. Only one branch of the conditional expression is evaluated in CBN, based on what the boolean expression evaluates to. Thus it is tempting to treat conditional expressions as functions, and use lifted function application to lift them. However, conditional expressions in a variability-aware setting are more subtle.

For example, assume we are lifting the following expression (with values from Example~\ref{ex:var}):

\[\kwif~(\kwiszero~\codevar{x})~\kwthen~(\codevar{x} + \codevar{z})~\kwelse~(\codevar{z}~/~\codevar{x}) \] 

If we are to treat this expression as a ternary function and apply its lifted version to the three arguments using Currying, we would violate the semantics of the original expression and throw a division-by-zero exception. The reason is that although CBN (lazy evaluation) evaluates expressions only when needed, both $(\codevar{x} + \codevar{z})$ and $(\codevar{z}~/~\codevar{x})$ actually need to be evaluated because $\kwiszero~\codevar{x}$ evaluates to $\kwtrue$~in configuration $\neg \FA \land \neg \FB$ and to $\kwfalse$~in the rest of the configurations. The results of the addition and division mathematical expressions will then be filtered based on the $\kwtrue$~and $\kwfalse$~configurations. In the general case, this is not optimal because some computation results will be thrown away, but even worse, in this particular case, it also results in a faulty behavior violating the intended semantics (division-by-zero).

To preserve the semantics of conditional expressions, we need to filter the expressions of the $\kwthen$~and $\kwelse$~ branches \emph{before} they are evaluated as follows:

\begin{prooftree}
	\AxiomC{$\lift{t_1} \eval \{(\kwtrue,x), (\kwfalse,y)\} $}
	\RightLabel{lifted-if}
	\UnaryInfC{$\lift{(\kwif~t_1~\kwthen~t_2~\kwelse~t_3)} \eval \semantics{\indexProd{\lift{t_2}}{x}} \cup \semantics{\indexProd{\lift{t_3}}{y}}$}	
\end{prooftree}

\begin{example}[Conditional Expressions]
	According to the lifted-if inference rule, this expression evaluates as follows:
	
	\begin{tabular}{l l}
		& $\lift{(\kwif~(\kwiszero~\codevar{x})~\kwthen~(\codevar{x} + \codevar{z})~\kwelse~(\codevar{z}~/~\codevar{x}))}$ \\
		= & $\lift{\kwif}~(\lift{\kwiszero}~\codevar{x})~\lift{\kwthen}~(\codevar{x} \lift{(+)} \codevar{z})~\lift{\kwelse}~(\codevar{z}~\lift{(/)}~\codevar{x})$ \\
		= & $\lift{\kwif}~\{(\kwiszero~\lit{1}, \FA),(\kwiszero~\lit{2}, \neg \FA \land \FB),(\kwiszero~\lit{0}, \neg \FA \land \neg \FB)\}$ \\
		& $\lift{\kwthen}~(\codevar{x} \lift{(+)} \codevar{z})~\lift{\kwelse}~(\codevar{z}~\lift{(/)}~\codevar{x})$ \\
		= & $\lift{\kwif}~\{(\kwfalse, \FA \lor (\neg \FA \land \FB)), (\kwtrue, \neg \FA \land \neg \FB)\}~\lift{\kwthen}~(\codevar{x} \lift{(+)} \codevar{z})~\lift{\kwelse}~(\codevar{z}~\lift{(/)}~\codevar{x})$ \\
		= & $\indexProd{(\codevar{x} \lift{(+)} \codevar{z})}{\neg \FA \land \neg \FB}~\cup~\indexProd{(\codevar{z}~\lift{(/)}~\codevar{x})}{\FA \lor (\neg \FA \land \FB)}$ \\
		= & $(\{(\lit{0}, \neg \FA \land \neg \FB)\} \lift{(+)} \{(\lit{19}, \neg \FA \land \neg \FB)\})~\cup~(\{\lit{19}, \FA \lor (\neg \FA \land \FB)\}~\lift{(/)}~\{(\lit{1}, \FA), (\lit{2}, \neg \FA \land \FB)\}$ \\
		= & $\{(\lit{19}, \neg \FA \land \neg \FB)\})~\cup~\{(\lit{19}, \FA), (\lit{9}, \neg \FA \land \FB)\}$ \\
		= & $\{(\lit{19}, \neg \FA \land \neg \FB), (\lit{19}, \FA), (\lit{9}, \neg \FA \land \FB)\}$ \\
	\end{tabular}
\end{example}

Operationally, to rewrite an arbitrary conditional expression we need two auxiliary rewrite operations: \term{lifted expression partitioning}, and \term{recursive expression restriction}. Given a lifted expression $e$ of type $\lift{T}$, a predicate $p$ of type $T \to \tbool$, the \term{partition} operator fully evaluates $e$ to lifted value $v$, then returns a pair of presence conditions $(pc_1,pc_2)$, where $pc_1$ is the disjunction of presence conditions of atomic values of $v$ satisfying the predicate $p$, and $pc_2$ is the disjunction of presence conditions of the other atomic values. 


The \term{restrict} rewriting operator takes a syntactic expression $e$ and returns a syntactic lambda expression taking a presence condition $pc$ as argument, and a body of $e$ where each subterm $t$ is replaced with $\indexProd{t}{pc}$. Restricting all the terms of an expression to a set of products results in restricting the whole expression to that set. A conditional expression is then deeply lifted as follows:

\begin{tabular}{l c l}
	&& \\
$\lift{(\kwif~c~\kwthen~t_1~\kwthen~t_2)}$ & $\rewrite$ &
	 $(pc_1, pc_2) = partition(c, \lambda b. b)$ \\
& & $((restrict~t_1)~pc_1) \cup ((restrict~t_2)~pc_2)$ \\
&& \\
\end{tabular}
\subsection{Lifting Pattern Matching}

\begin{figure*}[t]
\includegraphics[width=0.9\textwidth]{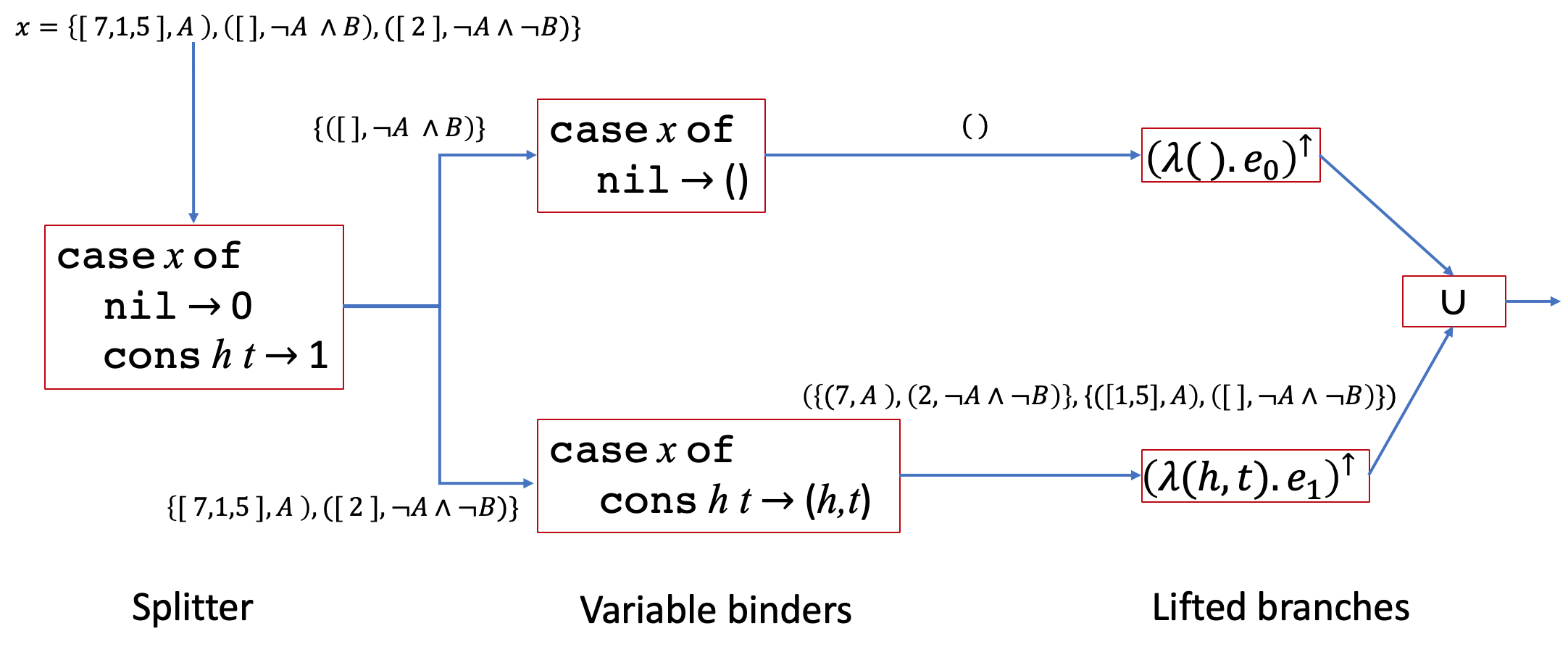}
\vspace{-0.1in}
\caption{The pipeline of the lifted case expression, applied to an example of a lifted list of integers.}
\vspace{-0.1in}
\label{fig:liftedCase}	
\end{figure*}

Pattern matching generalizes conditional expressions to an arbitrary number of potentially overlapping conditions, each with a branching alternative. Moreover, a pattern might include variables that are bound to values as a part of the matching process. For example, the \pcfp~program in Fig.~\ref{fig:tokenCount} matches the list expression $x$ against two patterns: $\kwnil$ for empty lists, and $\kwcons~h~t$~for non-empty ones. The second pattern binds the variables $h$ to the head of $x$, and $t$ to its tail.

In addition to splitting the lifted input across the different branching alternatives based on the matched patterns, lifted pattern matching needs to address two additional aspects: overlapping conditions and pattern variable binding. Since the conditions in a $\kwcase$ expression might overlap (i.e., more than one can evaluate to true), a priority mechanism is usually specified as a part of the semantics of the language. In \pcfp~(see Fig.~\ref{fig:semantics}), patterns are evaluated in order, and the branching alternative of the first matching pattern is the one evaluated. We need to preserve the same semantics when lifting a $\kwcase$~expression. In addition, variables of the matching pattern need to be bound to values from the expression we are matching against. Those variables are in the scope of the corresponding branch expression, which is to be deep-lifted. 

We lift $\kwcase$~expressions into $\lift{\kwcase}$ using a three-stage pipeline: splitting, binding and evaluation. This pipeline is demonstrated in Fig.~\ref{fig:liftedCase} on an example of matching against a list expression. The first stage (the splitter) is a $\kwcase$ expression mapping each element of the lifted input to the index of the matched pattern (0,1,...), The output of the first stage is a partition of the input lifted value. In the example, the input is partitioned into empty lists (matching the first pattern), and non-empty ones (matching the second). The first stage generates the following partitioning expression:

\begin{tabular}{l c l}
	&& \\
	$\lift{\kwcase~x~\kwof~p_0 \to t_0,...,p_n \to t_n}$ & $\rewrite$ & $\kwcase~x~\kwof~p_0 \to 0, ..., p_n \to n$\\
	&& \\
\end{tabular}

The second stage binds pattern variables to tuples of lifted values. The first pattern in the example does not bind any variables, so the output of the first binder is an empty tuple. The second pattern binds variable $h$ to ${(7,\FA), (2,\neg \FA \land \neg \FB)}$, and binds $t$ to ${([1,5], \FA), ([], \neg \FA \land \neg \FB)}$. Those are respectively the heads and tails of the matched elements of the input lifted value. The second stage generates the following variable binding expressions (where $vars(p_i)$ is the syntactic tuple $(v_{i,0}, ..., v_{i,k})$ of variables in pattern $p$):

\begin{tabular}{l c l}
	&& \\
	$\lift{\kwcase~x~\kwof~p_0 \to t_0,...,p_n \to t_n}$ & $\rewrite$ &
	$\kwcase~x~\kwof~p_0 \to vars(p_0)$ \\
	& & ... \\
	& & $\kwcase~x~\kwof~p_n \to vars(p_n)$ \\
	&& \\
\end{tabular}
	
The third stage is the branches of the original $\kwcase$~expression, each transformed into a lifted body of a lambda taking the tuple of bound values from stage two as input. The indices calculated in stage one are used to map atomic values to their matched lifted branches. The outputs of stage three are combined into a single lifted value, which is a straightforward set union operation. Each of the $t_k$ branching expressions of the $\kwcase$ expression is rewritten as follows:

\begin{tabular}{l c l}
	&& \\
	$\lift{t_k}$ & $\rewrite$ &
	$\lambda(vars(p_k)).restrict(t_k)$ \\
	&& \\
\end{tabular}

Recall that \emph{restrict} generates a lambda expression taking a presence condition as argument. The outputs from stage two together with presence conditions are passed to each of the expressions generated by stage three, and the union of the results is the result of the lifted $\kwcase$ expression.

\subsection{Deep Data}
\label{sec:deepData}

\begin{figure*}
	\begin{subfigure}[b]{0.39\textwidth}
		\includegraphics[width=0.7\textwidth]{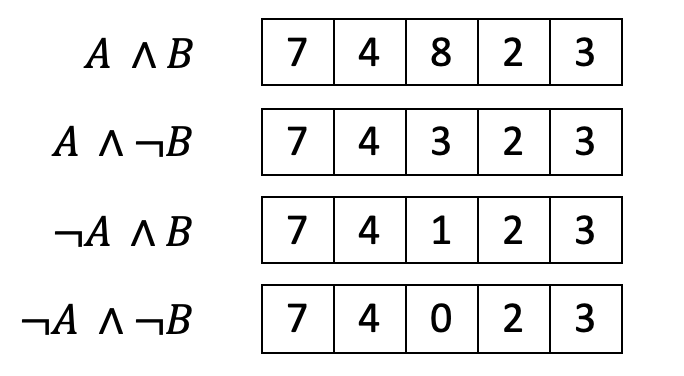}
		\caption{Shallow-lifted List.}
		\label{fig:shallowList}
	\end{subfigure}
	\hfill
	\begin{subfigure}[b]{0.6\textwidth}
		\includegraphics[width=0.7\textwidth]{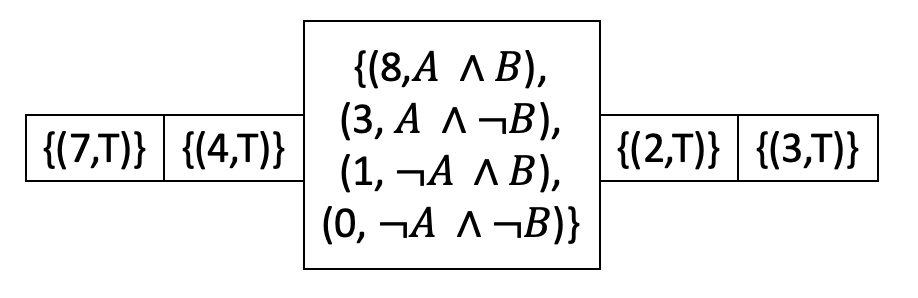}
		\caption{Deep-lifted List.}
		\label{fig:deepList}
	\end{subfigure}
\caption{Difference between (a) deep-lifted and (b) shallow-lifted lists.}
\label{fig:liftedLists}
\end{figure*}

Monomorphic data types are lifted into sets of value-presence condition pairs. However, polymorphic data types (e.g., lists and pairs) carry values at multiple levels. For example, $\kw{[nat]}$ is a data structure at the outer level, carrying natural number values at the inner level. A shallow-lifted list (Fig.~\ref{fig:shallowList}) attaches a presence condition to the list data structure, keeping the inner values with their original type ($\tnat$ in this example). A deep-lifted list on the other hand (Fig.~\ref{fig:deepList}) keeps the original data structure as is, while lifting the inner values. 

Under a deep-lifting framework, polymorphic data types (list and pairs) are lifted as follows:
\[ \lift{\tlist} \rewrite [\lift{T}]\]
\[\lift{(T_1, T_2)} \rewrite (\lift{T_1}, \lift{T_2})\]

With the exception of $\lift{\kwisnil}$, using the same implementation of their unlifted counterparts, lifted list and pair operations have the following type signatures:
\[ \lift{\kwcons} : \lift{T} \to [\lift{T}] \to [\lift{T}]\]
\[ \lift{\kwhead} : [\lift{T}] \to \lift{T} \]
\[ \lift{\kwtail} : [\lift{T}] \to [\lift{T}] \]
\[ \lift{\kwisnil} : [\lift{T}] \to \lift{\tbool} \]
\[ \lift{\kwfst} : (\lift{T_1}, \lift{T_2}) \to \lift{T_1} \]
\[ \lift{\kwsnd} : (\lift{T_1}, \lift{T_2}) \to \lift{T_2} \]

Because $\lift{\kwisnil}$ is expected to return a lifted Boolean, we cannot use its unlifted version as is. Instead, it needs to pair the result of calling $\kwisnil$ with \TT~as a presence condition (because of the total coverage invariant) as follows:
\[ \lift{\kwisnil}~x = \{(\kwisnil~x,\TT)\}\]
Deep lifting of data structures has the advantage of minimizing data redundancy, and maximizing sharing. The example in Fig.~\ref{fig:liftedLists} shows a variability-aware list of five elements, where the third element varies across different product variants, while all the other elements are the same across all variants. A shallow-lifting would result in four lists with four out of five positions carrying the same value. Even if the language compiler and run-time system are smart enough to share the memory where common elements are stored, when the shallow-lifted list is traversed the common elements will still be read multiple times.

The disadvantage of deep lifting of data structures on the other hand is that they need special case handling of their type signatures. The lifting framework assumes that any type $\T$~is lifted to type $\lift{\T}$. For example, a lifted list of natural numbers is expected to have the $\lift{\kw{[nat]}}$ type. A deep-lifted list of naturals would have the type $\kw{[\lift{nat}]}$ instead. In order to support deep lifting of data structures, we treat polymorphic types differently during rewriting. This requires maintaining type information of terms and expressions during the rewriting process.

\section{Correctness of Lifted Function Application}
\label{sec:correctness}

\newcommand{\e}{{e}}
\renewcommand{\c}{{c}}
\renewcommand{\v}{{v}}
\renewcommand{\a}{{a}}
\renewcommand{\b}{{b}}
\renewcommand{\d}{{d}}

\newcommand{\up}[1]{#1^\uparrow}
\newcommand{\fup}{\up{\f}}
\newcommand{\xup}{\up{\x}}
\newcommand{\vup}{\up{\v}}

\newcommand{\vp} {{v'}}

\renewcommand{\C}{{C}}

\newcommand{\fpc}{{fpc}}
\newcommand{\xpc}{{xpc}}
\newcommand{\vpc}{{vpc}}

Deep rewriting rules preserve the semantics of their respective language constructs by design. The underlying primitives of the lifting framework are lifted values (with their invariants) and lifted function application. In this section, we specify the correctness criteria of lifted programs, and prove that lifted function application preserves the disjointness and total coverage invariants of lifted values, and the correctness criteria of lifting.

\BD[\term{Correctness of Lifting}]
\label{def:correctness}
Given a product line $\L$, a \pcfp~function $\f$ and a configuration $\config$, $\fup$ is a correct lifting of $\f$ if and only if indexing $\fup$ applied to $\L$ by $\config$ is equal to the result of applying $\f$ to the product generated from $\L$ using configuration $\config$. Formally, $\indexProd{\fup(\L)}{\config} = \f(\indexProd{\L}{\config})$.
\ED

This is summarized by the following commuting diagram:

\begin{figure}[h]
\begin{tikzcd}
	\L \arrow{r}{\lift{\f}} \arrow{d}{\indexProd{}{\config}} & \lift{R} \arrow{d}{\indexProd{}{\config}} \\
	\indexProd{\L}{\config} \arrow{r}{\f} & R
\end{tikzcd}
\end{figure}

\subsection{Application Preserves Disjointness}

\newtheorem{th1}[theorem]{Theorem}{\bfseries}{\itshape}
\begin{th1}
\label{th:disjointness}
If both $\fup$ and $\xup$ satisfy the disjointness invariant, then the result of $(\apply~\fup~\xup)$ also satisfies disjointness.
\end{th1}

\begin{proof}

By contradiction:

Assume that both $\fup$ and $\xup$ satisfy the disjointness invariant while the result $\y = (\apply~\fup~\xup)$ does not. Then there exists at least one pair of presence conditions in $\y$ ($\pc{i}$ and $\pc{j}$) where $\sat(\pc{i} \wedge \pc{j}) = \true$. 

Each of $\pc{i}$ and $\pc{j}$ is a conjunction of presence conditions from $\fup$ and $\xup$ (the $\apply$ inference rule), then for arbitrary $\a$, $\b$, $\c$ and $\d$:
$$\begin{array}{lcl}
\pc{i} & = & \fpc_a \wedge xpc_b\\
\pc{j} & =  & \fpc_c \wedge \xpc_d
\end{array}$$

Then $\pc{i} \wedge \pc{j} = \fpc_a \wedge \xpc_b \wedge \fpc_c \wedge \xpc_d$.

For this conjunction to be satisfiable, all pair-wise conjunctions of its terms need to be satisfiable. This includes both $\sat(\fpc_a \wedge \fpc_c)$ and $\sat(\xpc_b \wedge \xpc_d)$, which contradicts the assumption that $\fup$ and $\xup$ both satisfy disjointness. Then $y = (\apply~\fup~\xup)$ must satisfy disjointness too.
\end{proof}

\subsection{Application Preserves Full Coverage}

\newtheorem{th2}[theorem]{Theorem}{\bfseries}{\itshape}
\begin{th2}
	\label{th:fullCoverage}
	If both $\fup$ and $\xup$ satisfy the full coverage invariant, then the result of $(\apply~\fup~\xup)$ also satisfies full coverage.
\end{th2}

\begin{proof}
	
	Given an arbitrary product configuration $\config \in \featmodel$, there exists $(f',fpc) \in \fup$ and $(x',xpc) \in \xup$ where $\indexProd{\fup}{\config} = f'$ and $\indexProd{\xup}{\config} = x'$ (since both $\fup$ and $\xup$ satisfy the full-coverage invariant).
	
	Then $(f'~x', fpc \land xpc) \in (\apply~\fup~\xup)$ (\code{apply} inference rule). Since $\config$ is an arbitrary configuration, then each configuration will be covered in the result of lifted application (i.e., lifted application preserves full coverage).
	
%
%
%
%
%
	
\end{proof}

\subsection{Application Preserves Correctness}

\newtheorem{th3}[theorem]{Theorem}{\bfseries}{\itshape}
\begin{th3}
If $\fup$ is a lifting of function $\f$, and $\xup$ is a lifting of value $\x$, then the result of~$(\apply~\fup~\xup)$ satisfies the correctness criteria (Definition~\ref{def:correctness}) with respect to $\f~\x$.
\end{th3}

\begin{proof}
\par Given $\vup = (\apply~\fup~\xup)$, we need to prove that given a configuration $\config$, $\indexProd{\vup}{\config} = \f(\indexProd{\xup}{\config})$.

\par According to the \code{lifted-type} and \code{apply} inference rules: 
	\[ \vup = \{(\f'(\x'),\pc{})~|~(\f', \fpc) \in \fup, (\x', \xpc) \in \xup, \pc{} = \fpc \wedge \xpc, \sat(\pc{})\} \]
	
\par Let \[ \c = \bigwedge_{\e \in F}
	\begin{cases}
			\e 		& \text{if}~\e \in \config \\
			\neg \e	& \text{otherwise}
	\end{cases}
	\]
This is the propositional formula representation of the configuration $\config$, given a feature set $\F$ (Definition~\ref{def:featconfig}).

\par Since $\apply$ preserves the full coverage invariant (Theorem~\ref{th:fullCoverage}), $\v = \indexProd{\vup}{\config}$ exists, such that $(\v, \vpc) \in \vup, \sat(\vpc \wedge \c)$.
\par Then $\v =  \indexProd{\vup}{\config} = \f(\indexProd{\vup}{\config})$ (definition of lifted value indexing).

\end{proof}

\section{Implementation}
\label{sec:impl}

We implemented both shallow and deep lifting in Haskell\footnote{\url{https://github.com/ramyshahin/ProductLineAnalysis}}, taking the subset of Haskell corresponding to \pcfp~as input. Our lifting framework is comprised of two components: a library and a rewriting engine. The library includes a set of variability-aware polymorphic types, primitive operations, and helper functions used in the deep lifting of some Haskell constructs (e.g., conditional and pattern matching expressions). The Deep Rewriter engine is a source-to-source transformer that takes a program in our supported subset of Haskell, and generates its deep-lifted version. The overall architecture of our lifting framework (together with benchmarking and preprocessing components) is outlined in Fig.~\ref{fig:arch}.

\subsection{Lifted Types}

The target language of our lifting engine is full-fledged Haskell, not just the \pcfp-compliant subset. 
Fig.~\ref{fig:library} shows snippets of type and function definitions from the Haskell variability library. We define a polymorphic variability-aware type \code{Var} as a list of pairs of values and presence conditions. We implement presence conditions as Binary Decision Diagrams (BDDs) using the CUDD library~\cite{Somenzi:1998}. Singleton \code{Var} values can be constructed using the \code{mkVar} function. The \code{mkVarT} function creates a singleton \code{Var} value with the \TT~presence condition (\code{ttPC}).

Shallow-lifted functions are applied to arguments using the \code{apply} function, which implements the semantics of the \code{apply} inference rule. The library includes variants of \code{apply} that support different arities. In addition, the \code{Var} type instantiates the Haskell \code{Applicative} type class, so functions and operators available for that type class can be readily used on \code{Var} values.

Our implementation of \code{apply} checks presence condition conjuncts for satisfiability \emph{before} applying the function to its argument. This way, if the conjunct is not satisfiable and the element is to be filtered out, we do not waste computational resources calculating a value that will be eliminated anyway. This is particularly important for computationally expensive functions.

\begin{figure*}[t]
\scalebox{0.8}{}
\begin{mbox}{}
\centering
\begin{lstlisting}[language=Haskell,basicstyle=\small]
type    Val a = (a, PresenceCondition)
newtype Var t = Var [Val t]

mkVar :: PresenceCondition -> t -> Var t

mkVarT = mkVar ttPC

apply :: Var (a -> b) -> Var a -> Var b

instance Applicative Var where
pure  = mkVarT
(<*>) = apply
\end{lstlisting}
\end{mbox}
\caption{Snippets of type and function definitions from the Haskell variability library.}
\label{fig:library}
\end{figure*}

\subsection{Deep Rewriter}

For deep lifting, we rewrite a Haskell program (in the \pcfp~subset of Haskell) into a semantically equivalent variability-aware Haskell program. Our program rewriting engine is based on the haskell-tools\footnote{\url{https://github.com/haskell-tools}} library, which provides a programmatic interface to the Haskell Abstract Syntax Tree (AST). We traverse the AST of the input program, applying the rewrite rules defined in Sec.~\ref{sec:deep} to each syntactic construct that we support. An error is reported if the input program includes any unsupported syntactic constructs.

Because we use deep lifting polymorphic data types as well (Sec.~\ref{sec:deepData}), we need to keep track of term and expression types to make sure that the generated program conforms to type rules. Our implementation does not support lifting user-defined types at this point, but we expect the effort needed for that to be minimal. In general, lifted user-defined product types would use the specialized infrastructure we currently have for lifting pairs, and lifted user-defined sum types would use the infrastructure we have for lifting lists.

The program being rewritten might be using definitions exported by other modules. We provide a command-line option for specifying whether any module imports in the input program should be replaced by imports of deep-lifted modules instead. This allows for deep-lifting a code-base incrementally, one module at a time. If there is a dependency on a module that has not been deep-lifted, calls to functions from that module are shallow-lifted. 

\subsection{Laziness, Conditionals and Pattern Matching}

Our theoretical treatment assumed Call-By-Name (CBN) semantics, where expressions in a control-flow branch are evaluated only when the branch is taken. In a variability-aware context, this is more complicated because multiple branches might be taken (for different subsets of lifted values), instead of only one. Filtering out values with unsatisfiable presence conditions before those values are computed ensures that expressions in a branch will only be applied to values where this branch is to be taken.

Haskell is a non-strict language (i.e., expression evaluation starts at the root of the expression AST, not the leaves), and  different Haskell compilers and run-time systems implement non-strictness differently. Our Haskell implementation exhibited behaviors different from those in our theoretical model. In particular, some branches of conditional and pattern matching expressions were overactive (i.e., they were taken in cases where they should not). In some cases, this resulted in infinite loops when the overactive branch has a recursive call.

To mitigate these erroneous behaviors, we extended the implementation of lifted control-flow constructs (conditionals and pattern matching) with explicit context passing to branches. A \term{context} is the presence condition of values where a branch needs to be taken. Before evaluating expressions in a branch, values in those expressions are first \emph{restricted} to the presence condition of the branch. Restriction here means removing atomic elements of those values that are outside the set of configurations defined by the branch presence condition. This ensures that even if a branch is overactive, its expressions will not be evaluated because their arguments will be turned into empty lifted values by restriction.

The lifted $\kwif$ and $\kwcase$ expressions evaluate their guards first, and pass the presence conditions of guard values matching each of the branches to the respective branch. The signatures of lifted $\kwif$ and lifted $\kwcase$ in our implementation are as follows:

\begin{lstlisting}[language=Haskell,basicstyle=\small]
liftedIf   :: Var Bool -> (PresenceCondition -> Var t) 
           -> (PresenceCondition -> Var t) -> Var t
liftedCase :: Var a -> (a -> Int) 
           -> [PresenceCondition -> Var a -> Var b] -> Var b

\end{lstlisting}
\noindent
where the second parameter of \code{liftedCase} is the splitter function, and the third parameter is a list of alternatives taking a \code{PresenceCondition} context as argument.

\section{Evaluation}
\label{sec:evaluation}

Brute-force analysis enumerates \emph{all} combinations of a product line, and applies the original analysis to each individually. The number of combinations is exponential in the number of features. The goal of variability-aware lifting of analyses is to improve over that worst-case exponential time. The two approaches to lifting presented in this paper come with a trade-off between leveraging sharing opportunities across variants on one hand, and the associated overhead of maintaining presence conditions on the other. 

In most cases, the piece of software to be analyzed only covers a subset of all possible combinations. We call that the number of \term{effective combinations} (Def.~\ref{def:effComb}). Shallow lifting enumerates the input vectors that correspond only to \emph{disjoint sets} of effective combinations. Taking the original analysis as a black-box, the shallow-lifted analysis only needs to track variability (in terms of presence conditions) of inputs and outputs. This is light-weight in terms of overhead, but that comes at the expense of missing all opportunities to share common intermediate values within the analysis.

Deep lifting, on the other hand, associates a presence condition with each value throughout the execution time of the lifted analysis. This way intermediate values that are common across different product variants are shared right away, reducing the overall number of computations required. However, each computation calculates both a value and a presence condition (a BDD in our implementation). BDD operations are exponential in the number of propositional variables (features) in the worst-case, but optimized BDD packages usually perform much better on moderately sized inputs.

Taking all the theoretical limitations and trade-offs into consideration, the goal of our evaluation is to answer two research questions:




\RQ{1}: How do shallow lifting and deep lifting scale with the growth in  the number of features (and effective combinations), compared to brute-force analysis?

\RQ{2}: At what point does deep lifting start outperforming brute-force analysis and shallow lifting?

\subsection {Evaluation Experiment Setup}

\begin{figure*}[t]
	\includegraphics[width=\textwidth]{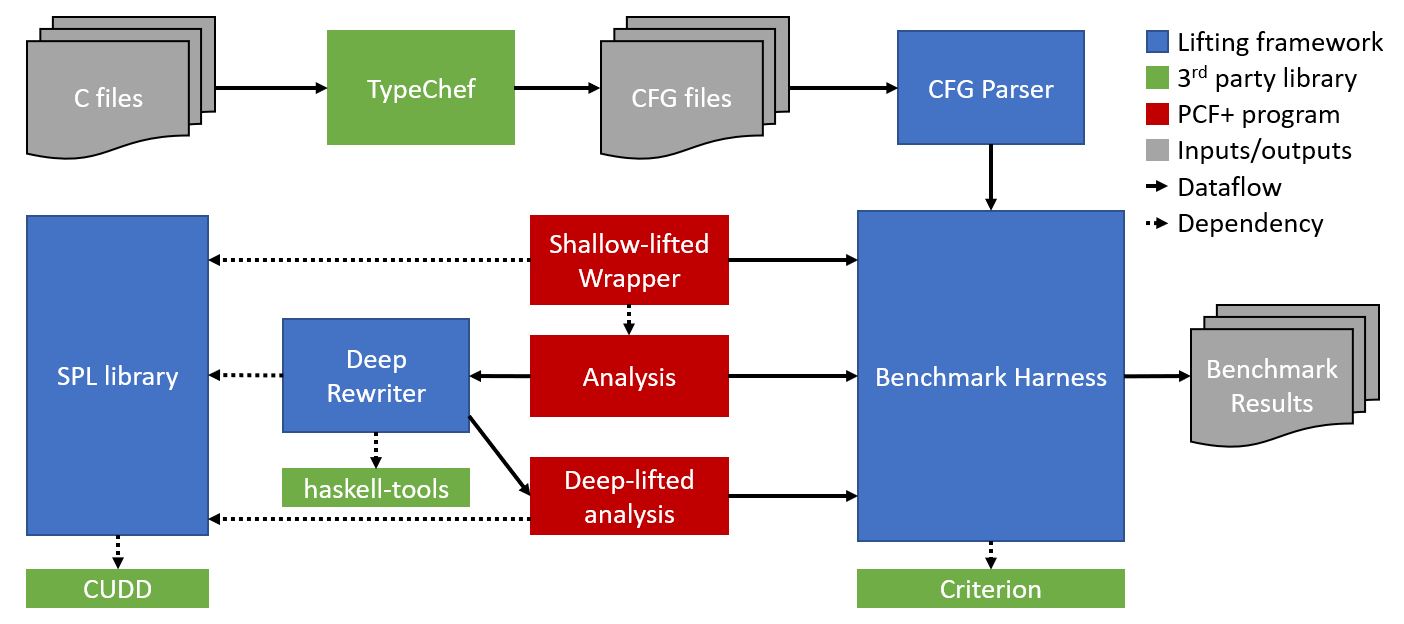}
	\caption{Architecture of the lifting framework, benchmarking components and C-file processing pipeline.}
	\label{fig:arch}
\end{figure*}

\begin{table*}[tbp]
\centering
\begin{tabularx}{\linewidth}{lX}
	\toprule
	Analysis & Description \\
	\midrule
	Case Termination & Checking whether each non-empty \texttt{case} in a C-language \texttt{switch} statement ends with a \texttt{break} statement. \\
	Dangling Switch & Checking if there is any dead-code (anything other than a declaration) between a \texttt{switch} statement and the first \texttt{case} or \texttt{default}. \\
	Function Return Checker & Checking whether each non-\texttt{void} C function has at least one \texttt{return} statement. \\
	Return Density & Calculates the average number of \texttt{return} statements per C function within a source file. \\
	Goto Density & Calculates the average number of \texttt{goto} statements per label within a C source file. \\
	Call Density & Calculates the average number of function calls per C function within a source file. \\
	\bottomrule
\end{tabularx}	
\caption{Description of program analyses used in the evaluation.}
\label{tbl:analyses}
\vspace{-0.1in}
\end{table*}

We implemented four C-language program analyses, described in Table~\ref{tbl:analyses}, in the \pcfp~subset of Haskell.  Each analysis takes a Control Flow Graph (CFG) of a C-language program as input, and either checks a property of the program or calculates a metric. Three of the analyses (Case Termination, Dangling Switch, and Function Return Checker) are adapted from \cite{Rhein:2018}, and the others were designed by us. 

The architecture of the benchmarking environment is outlined in Fig.~\ref{fig:arch}.
Each analysis is encapsulated in a brute-force harness, wrapped in a shallow-lifted version, and transformed using the Deep Rewriter into a deep-lifted version. The run-time performance of the three versions of each analysis is measured and compared using a benchmarking harness based on the Criterion\footnote{\url{http://www.serpentine.com/criterion/}} Haskell microbenchmarking library. Criterion runs each benchmark multiple times (20 times by default), and estimates the mean running time using linear regression.

We applied each benchmark to a collection of 504 C-language source files from BusyBox\footnote{\url{https://busybox.net/}} version 1.18.5. These source files varied in size, complexity, and number of effective combinations. Each source file was transformed into a variational CFG file using the TypeChef variability-aware framework~\cite{Kastner:2011}. We then parsed the CFG file into the input data structure of the analyses.
All experiments were performed on a Quad-core Intel Core i7-6700 processor running at 3.4GHZ, with 16GB RAM, running 64-bit Ubuntu Linux (kernel version 4.15), and using the Glasgow Haskell Compiler version 8.6.5.

\subsection{Evaluation Results}

\begin{figure*}
	
\begin{subfigure}[b]{\textwidth}
\centering
	\includegraphics[width=0.9\textwidth]{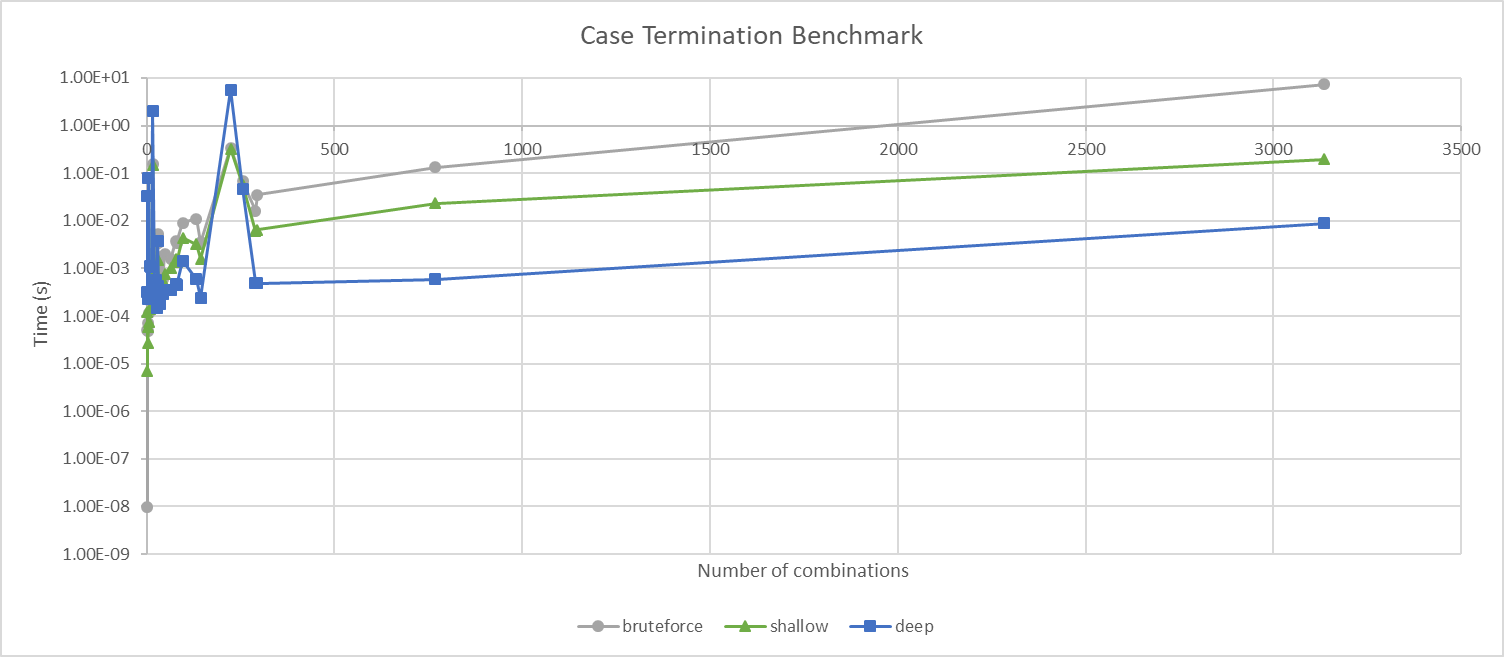}
	\caption{Case Termination benchmark.}
	\label{fig:caseTermination}
\end{subfigure}
\hfill
\begin{subfigure}[b]{\textwidth}
\centering
	\includegraphics[width=0.9\textwidth]{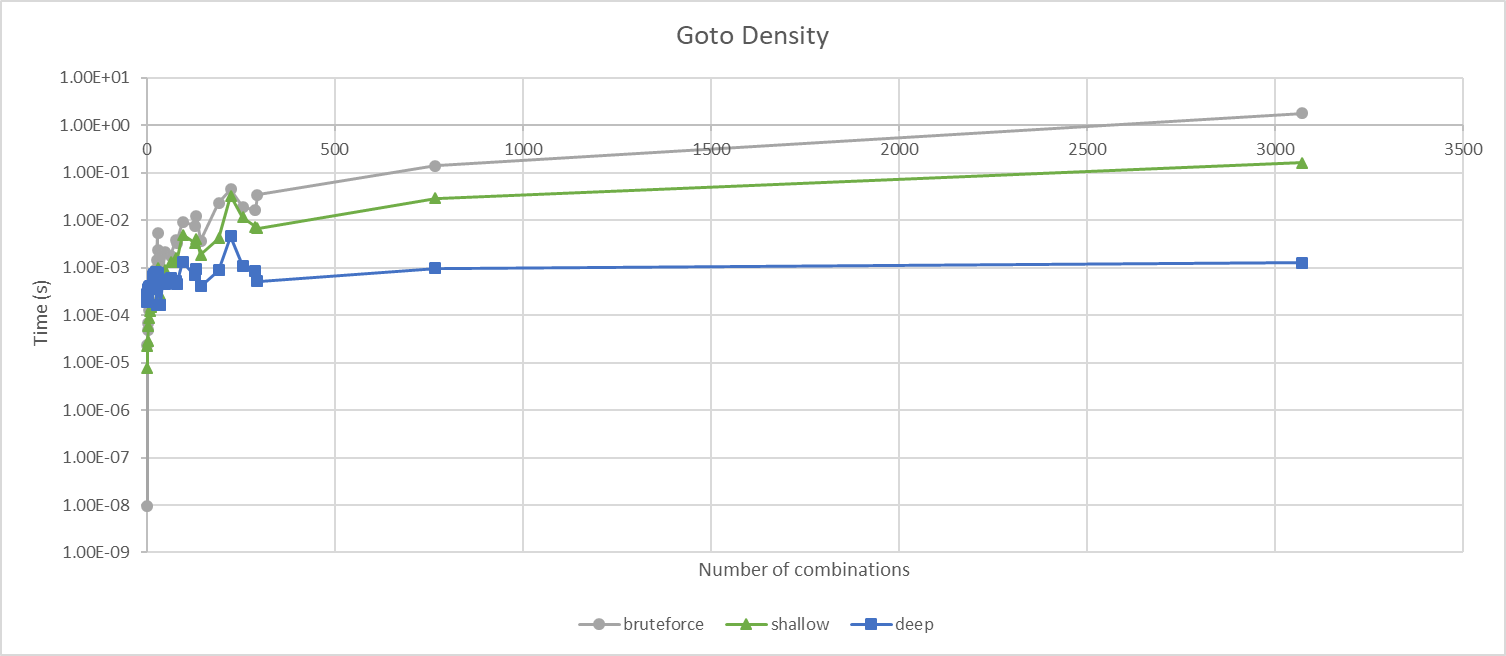}
	\caption{Goto Density benchmark.}
	\label{fig:goto}
\end{subfigure}
\begin{subfigure}[b]{\textwidth}
	\centering
	\includegraphics[width=0.9\textwidth]{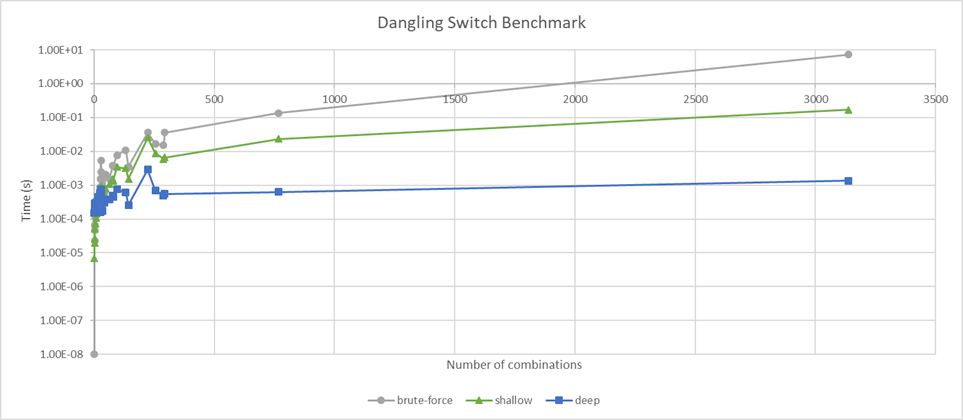}
	\caption{Dangling Switch benchmark.}
	\label{fig:danglingSwitch}
\end{subfigure}
\caption{Evaluation results on  (a) case termination, (b) goto density per label, and (c) dangling switch.  Horizontal axis is the number of effective product combinations in a source file, and vertical axis is average processing time of files with that number of configurations, in log-scale.}
\label{fig:benchmarks1}
\vspace{-0.1in}
\end{figure*}

\begin{figure*}
	
\begin{subfigure}[b]{\textwidth}
\centering
	\includegraphics[width=0.9\textwidth]{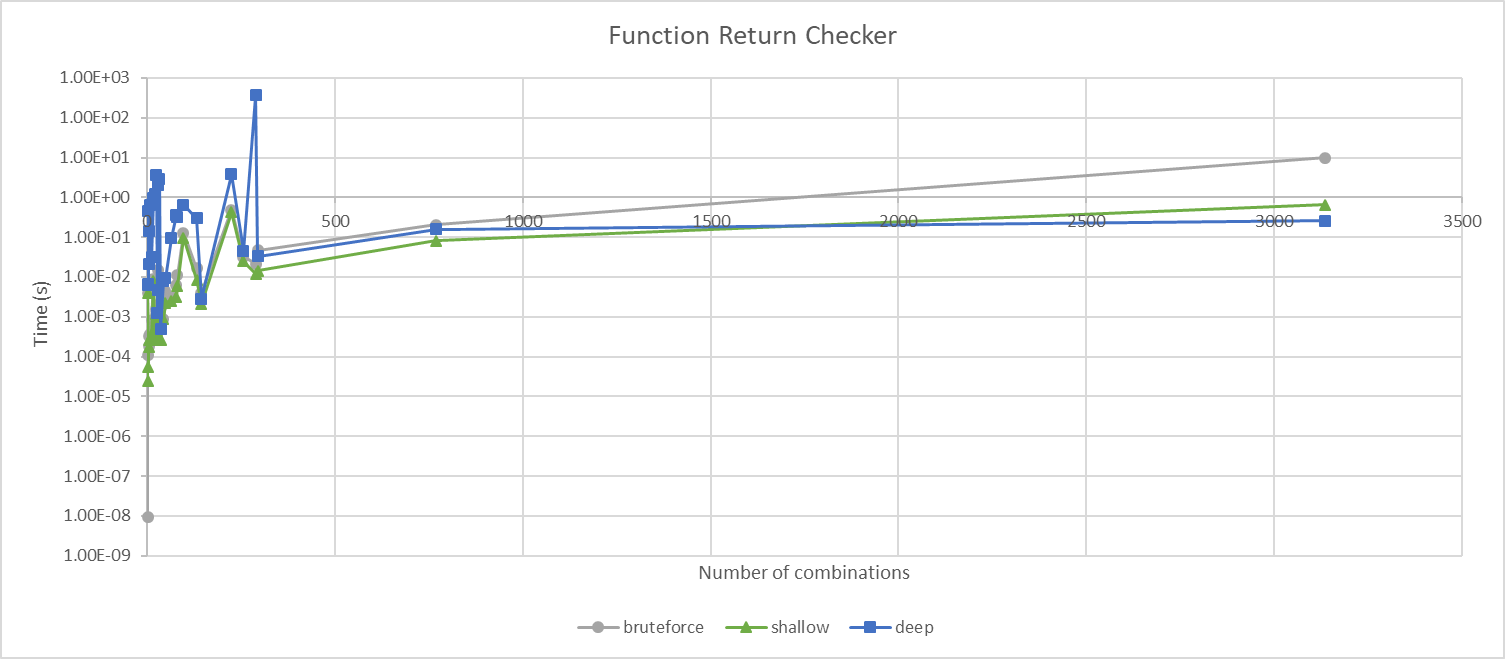}
	\caption{Function Return Checker benchmark.}
	\label{fig:return}
\end{subfigure}
\hfill
\begin{subfigure}[b]{\textwidth}
\centering
	\includegraphics[width=0.9\textwidth]{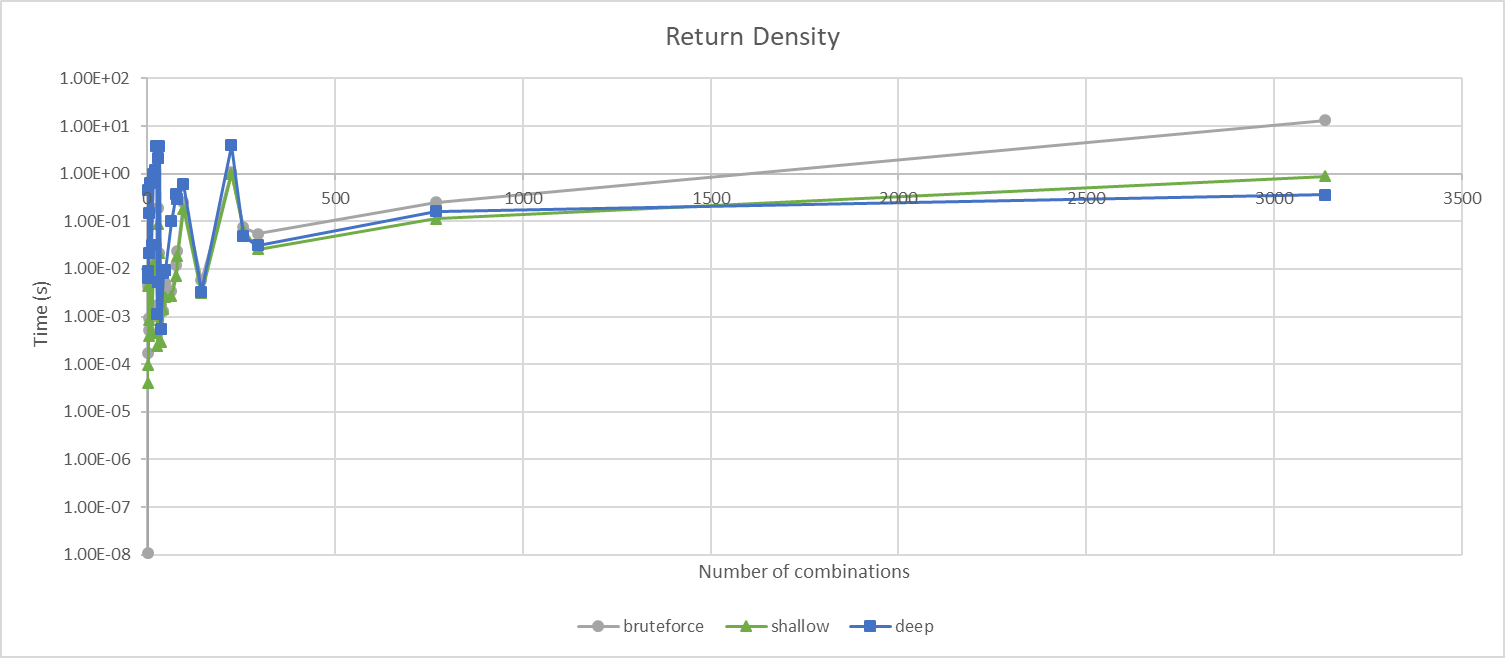}
	\caption{Return Density benchmark.}
	\label{fig:returnDensity}
\end{subfigure}
\hfill
\begin{subfigure}[b]{\textwidth}
	\centering
	\includegraphics[width=0.9\textwidth]{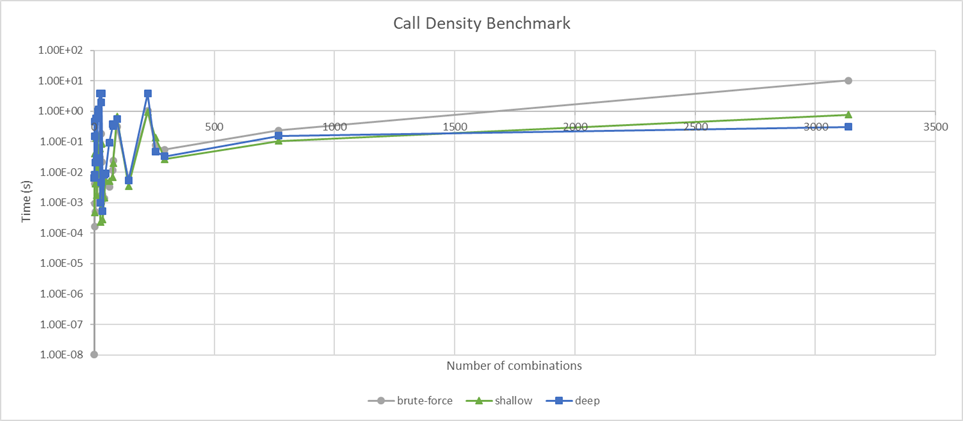}
	\caption{Call Density benchmark.}
	\label{fig:callDensity}
\end{subfigure}
\caption{Evaluation results on (a) function return checker, (b) return density per function, and (c) call density per function. Horizontal axis is the  number of effective product combinations in a source file, and vertical axis is average processing time of files with that number of configurations, in log-scale.}
\label{fig:benchmarks2}
\vspace{-0.1in}
\end{figure*}

Results of our evaluation experiments are plotted in Fig.~\ref{fig:benchmarks1} and \ref{fig:benchmarks2}.  We group the input source files based on their number of effective combinations, calculate the average run time for each group, and plot that against the number of effective combinations. The vertical axis of each of the six graphs is running time in seconds in log-scale.

For a very small number of combinations, brute-force outperforms shallow lifting by more than an order of magnitude and outperforms deep lifting by two to three orders of magnitude in each of the benchmarks. However, as the number of effective combinations increases, shallow lifting starts outperforming brute-force. The point at which deep lifting starts outperforming both brute-force and shallow lifting varies from one benchmark to another. In the Goto Density and Dangling Switch benchmarks, deep lifting takes over as the number of combinations go beyond 50. In Case Termination, deep lifting takes over at around 300 combinations. At 640 combinations, deep lifting is an order of magnitude faster than shallow lifting, and two orders of magnitude faster than brute-force in both benchmarks. However, in the Function Return Checker, Return Density, and Call Density benchmarks the performance of deep lifting fluctuates in the range from 10 to 300 combinations, and only slightly outperforms shallow lifting at around 3100 combinations (while outperforming brute-force by almost two orders of magnitude).

The input C files also vary in size and complexity, and this is likely the reason why we do not see a steady exponential growth in run time for brute-force analysis. The benchmark analyses also vary in terms of computational complexity, which explains the difference in performance patterns across them. 
For \RQ{1}, although the overhead of deep lifting is evident throughout the range of effective combinations, this overhead grows slower than the exponential costs of brute-force and shallow lifting. The answer to \RQ{2} varies across benchmarks, but overall we consistently see deep-lifting outperforming brute-force and shallow lifting as the number of effective combinations is in the hundreds.

\subsection{Threats to Validity}

Our benchmarks are all implementations of syntactic or control-flow analyses. 
Our framework is not specific to syntax or control flow analysis, but expanding to other categories of analysis requires infrastructure analogous to the TypeChef framework for variability-aware parsing and CFG generation. Also we cannot use existing analysis libraries directly because they typically use language features that we still do not support, so we had to write our own analyses.

As mentioned earlier, the C files being analyzed vary in size and complexity, and this skews the overall trend of results to some extent. This is more evident at the high end of combinations, as the number of source files per group is relatively small (sometimes only one), which negatively affects the generality of the results. However, since we care more about the overall trends of scalability as the number of combinations grows, we believe we can tolerate those skews. 

\section{Discussion}
\label{sec:discussion}

\vskip 0.1in
\noindent
{\bf Relaxing Total Coverage.}
The total coverage invariant states that a lifted value should cover the complete set of product variants defined by the feature model. 
Pragmatically, this over-constrains lifted values in some cases. For example, in Fig.~\ref{fig:annotative} the argument $b$ of type \tm{int} only exists when feature $\FA$ is not defined. Storing a representation of this argument by itself in a lifted variable would again violate the Full Coverage Invariant because it is not only defined in a proper subset of products.

This invariant can be relaxed using an \tm{Option} type (\code{Maybe} in Haskell) for the atomic elements of a lifted value. This way we can use \code{Nothing} as a valid value in products where the lifted value is not defined. This way the invariant is still strictly maintained, while allowing for incompleteness with respect to product space. Computing over \code{Nothing} values will still be a challenge though, as the correct behavior is most likely application dependent.

\vskip 0.1in
\noindent
{\bf  Optimization and Canonicalization.}
The bigger the size of a lifted value (in terms of the number of atom values within it), the longer it takes (in terms of processing time) to process it. Recall that when applying a lifted function to a single lifted argument, we compute a cross-product of atom values, and check the satisfiability of the conjunction of presence conditions of each pair in the cross-product. This becomes a major performance bottleneck in any lifted program. 

Of course, we cannot eliminate individual values from a lifted value. However, we can put a lifted value in \emph{canonical} form by combining common values. For example, if a lifted value has two atom values $(v_1, pc_1)$ and $(v_2, pc_2)$, and if $v_1$ and $v_2$ happen to be the same, then the two individual values can be collapsed into $(v_1, pc_1 \vee pc_2)$. Determining if $v_1$ and $v_2$ are the same can be done by checking their addresses in memory (a quick check) or by checking if the actual values (and recursively those of inner fields) are the same (a more costly check).


\vskip 0.1in
\noindent
{\bf Beyond SPLs.}
Lifting a TLC program to a variable domain (where variability is labeled by presence conditions) is just one example of reusing the same algorithm in multiple computational contexts at once. Other examples include \tm{probabilistic computing}, where a lifted value would have several atomic values each with a probability, and with the individual probabilities summing up to one. 
We can then directly map the different constructs specific to presence condition to those of probabilities. Assuming independent probabilities, conjunction of presence conditions is isomorphic to multiplication of probabilities. The disjointness and full coverage invariants are together isomorphic to the invariant that the sum of probabilities within a lifted value is always one.

\section{Related Work}
\label{sec:related}
Several software analyses have been re-implemented to be variability aware~\cite{Thum:2014}. For example, the TypeChef project~\cite{Kastner:2011,Kastner:2012} implements variability aware parsers~\cite{Kastner:2011} and type checkers~\cite{Kastner:2012} for the Java and C languages. The SuperC project~\cite{Gazzillo:2012} is another C language variability-aware parser. Some variability-aware control-flow and data-flow analyses have been implemented on top of TypeChef~\cite{Rhein:2018}. These are all lifted analyses designed and implemented from scratch, as opposed to our approach of lifting an analysis \emph{automatically}.

SPL\textsuperscript{Lift}~\cite{Bodden:2013} extends IFDS~\cite{Reps:1995} data flow analyses to product lines. Model checkers based on Featured Transition Systems~\cite{Classen:2013} check temporal properties of transition systems where transitions can be labeled by presence conditions. These are two examples of SPL analyses that use almost the same single-product analyses on a lifted data representation. Our approach lifts algorithms together with the data structures to which they are applied.

Lifting a language instead of lifting individual analyses has been attempted for Datalog~\cite{Shahin:2019, Shahin:2020}. In this work, Datalog facts are assigned presence conditions, and a Datalog engine is modified to interpret and compute presence conditions together with inferring new facts. Our lifted values are quite similar in the sense that we associate presence conditions to atomic values, and compute them in lifted function application. We extend that notion  in three ways. First, lifted values in our framework can hold multiple atomic values and presence conditions (subject to invariants). Second, and more importantly, we lift a Turing-complete language (\pcfp) capable of expressing arbitrary analyses not expressible in Datalog. Finally, because of the computational expressiveness of \pcfp, we did not need to modify a language interpreter. Instead, a source-to-source transformation of programs was sufficient.

Syntactic transformation systems have been suggested for lifting abstract interpretation analyses to SPLs~\cite{Midtgaard:2015}. In this line of work, a systematic approach to lifting abstract interpretation analyses is outlined, together with correctness proofs. However, while this approach is systematic, it is not automated, requiring manually designed abstractions.  In contrast, our approach is fully automated.

Empirical comparisons between brute force (product-based), variability-aware (family-based) and sampling techniques have also
been conducted~\cite{Apel:2013,Liebig:2013}. Variability-aware analyses have been shown to outperform sampling, while not giving up configuration completeness.

Automated lifting of type-based analyses~\cite{Chen:2014} to support annotative product lines is another attempt towards solving the SPL lifting problem. However, this approach only works if the analysis can be expressed as a type system. This restriction does not apply to our work.

Formalisms for developing variability-aware programs, e.g., the Choice Calculus~\cite{Erwig:2011}, have been introduced. Those formalisms introduce variability constructs on top of conventional single-valued, deterministic formalisms and  provide better language support for developing variability-aware analyses. However, to the best of our knowledge, automatic lifting of programs for those formalisms has not been attempted.

\section{Conclusion}
\label{sec:conclusion}
In this paper, we presented two approaches to automatic variability-aware lifting of software analyses written in a functional language (\pcfp). A light-weight, black-box approach (shallow lifting) wraps an analysis in its variability-aware version. The second approach (deep lifting) is an automatic program-rewriting mechanism that translates an analysis program into a semantically equivalent variability-aware analysis.

We presented and discussed the program rewriting rules for different language constructs. We also presented formal correctness criteria of lifted programs, and showed how our approach conforms to it. We implemented our lifting framework in Haskell, and evaluated it on a set of program analyses. Our evaluation results show that shallow lifting outperforms brute-force analysis on Software Product Lines with relatively small number configurations. As the number of configurations increases, deep lifting outperforms both brute-force and shallow lifting.

For future work, we plan to extend our approach to supporting language constructs beyond \pcfp. We also plan to work on improving the performance of lifted programs (lowering the overhead of presence condition computation). One more area we intend to explore is implementing and evaluating different policies for lifting data structures.




\section* {Acknowledgments}
We thank anonymous reviewers for their feedback, comments, and suggestions. This work was supported by General Motors and NSERC.

\bibliography{spl,pl}

\end{document}